\documentclass[aps,prb,twocolumn,showpacs,preprintnumbers,superscriptaddress]{revtex4}

\usepackage{amsmath}
\usepackage{amssymb}
\usepackage{graphicx}%
\usepackage{dcolumn}%
\usepackage{bm}%
\usepackage{color}%

\begin{document}
\title{Ab initio complex band structure of conjugated polymers: \\
       Effects of hydrid DFT and GW schemes}

\author{Andrea \surname{Ferretti}}
\email[corresponding author: ]{andrea.ferretti@unimore.it}
\affiliation{Centro S3, CNR--Istituto Nanoscienze, I-41125 Modena, Italy.}
\author{Giuseppe \surname{Mallia}}
\affiliation{Department of Chemistry, Imperial College London, London SW7 2AZ, UK.}
\author{Layla \surname{Martin-Samos}}
\affiliation{CNR -- IOM Democritos, I-34014 Trieste, Italy.}
\author{Giovanni~\surname{Bussi}}
\affiliation{Scuola Internazionale Superiore di Studi Avanzati, SISSA
via Bonomea 265, I-34136 Trieste, Italy.}
\affiliation{CNR -- IOM Democritos, I-34014 Trieste, Italy.}
\author{Alice \surname{Ruini}}
\affiliation{Dipartimento di Fisica, Universit\`a di Modena e Reggio Emilia, I-41125 Modena, Italy.}
\affiliation{Centro S3, CNR--Istituto Nanoscienze, I-41125 Modena, Italy.}
\author{Barbara \surname{Montanari}}
\affiliation{Computational Science and Engineering Department, 
             STFC Rutherford Appleton Laboratory, Oxfordshire OX11 0QX, UK.}
\author{Nicholas M. \surname{Harrison}}
\affiliation{Computational Science and Engineering Department, 
             STFC Daresbury Laboratory, Cheshire WA4 4AD, UK.}
\affiliation{Department of Chemistry, Imperial College London, London SW7 2AZ, UK.}

\date{\today}

\begin{abstract}
The non-resonant tunneling regime for charge transfer across nanojunctions
is critically dependent on the so-called $\beta$ parameter, governing the exponential
decay of the current as the length of the junction increases.
For periodic materials, this parameter can be theoretically evaluated by computing the 
complex band structure (CBS) -- or evanescent states -- of the material 
forming the tunneling junction.
In this work we present the calculation of the CBS for organic polymers
using a variety of computational schemes, including standard local, semilocal, and hybrid-exchange density functionals,
and many-body perturbation theory within the GW approximation.
We compare the description of localization and $\beta$ parameters
among the adopted methods and with experimental data.
We show that local and semilocal density functionals systematically 
underestimate the $\beta$ parameter, while hybrid-exchange schemes
partially correct for this discrepancy, resulting in a much better agreement with
GW calculations and experiments. Self-consistency effects and self-energy representation issues of the 
GW corrections are discussed together with the use of 
Wannier functions to interpolate the electronic band-structure.
\end{abstract}

\maketitle

\section{Introduction}
\label{sec:intro}

The fields of molecular electronics and charge transport through nanojunctions
have been deeply investigated in the past fifteen years.~\cite{agra+03prep,nitz-ratn03sci,floo+04sci}
At the experimental level many different techniques have been developed, including
those based on break junctions, nanostructured
and scanning probe layouts, or self-assembled monolayers.~\cite{agra+03prep,bour01book} 
Significant improvements in the accuracy with which these junctions are characterized have been achieved over
the years, {\it eg} to
address the I-V characteristics of single molecular junctions.
Moreover, a large experimental literature exists~\cite{holm+01jacs,chab+02jacs,choi+08sci} 
on non-resonant tunneling experiments, where 
it is possible to determine the 
exponential decay ($\beta_0$) of the current $I=I_0 \exp(-\beta_0 L)$ as a function of the length $L$ of 
the tunneling layer [such as {\it eg} a layer of organic self-assembled monolayer (SAM) connected
to metallic electrodes]. 
Even though the $\beta_0$ parameter depends~\cite{chab+02jacs,tomf-sank02prb,prod-car09prb,peng+09jpcc}
also on the detailed nature of the interface, it carries mostly information about the properties of the 
tunneling layer itself, which makes $\beta_0$ an interesting analysis and characterization tool.

\begin{figure}
   \includegraphics[clip, width=0.50\textwidth]{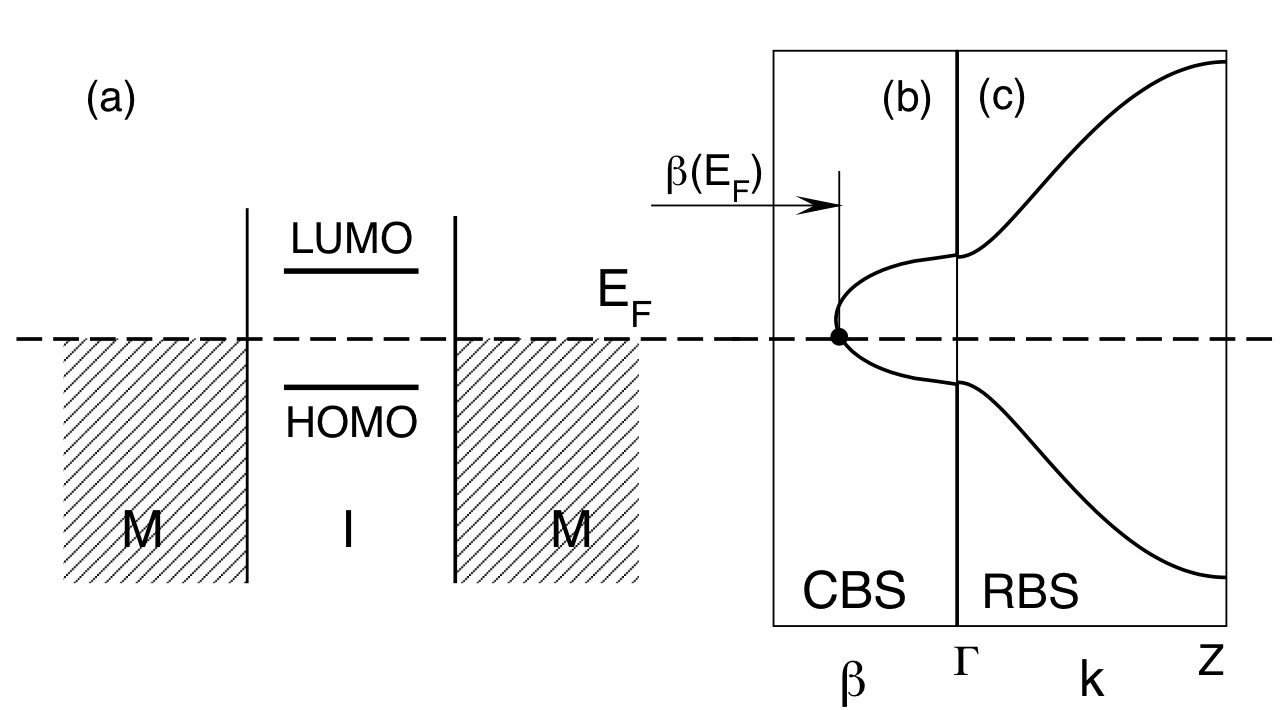}
   \caption{ \label{fig:cartoon}
        The non-resonant tunneling experiment.
        (a) Scheme of the alignment of electronic levels in a metal-insulator-metal (MIM)
        junction. The complex and real band-structure (CBS and RBS) corresponding to the (extended)
        insulator system are reported in (b) and (c) respectively. The computed value
        to be compared with experiments is highlighted as $\beta(E_F)$, $E_F$ being the
        Fermi energy of the MIM junction.}
\end{figure}

Experimentally, these measurements are performed using different setups, ranging from
metal-insulator-metal (MIM) junctions, as mentioned above, to 
the evaluation of kinetic constants of 
electro-transfer reactions (optically or electrochemically induced) 
in donor-bridge-acceptor molecular complexes.~\cite{oeve+87jacs,fink-hans92jacs,smal+95jpc,paul+93jpc,davi+98nat} 
In terms of systems, measurements have been performed on a number of cases ranging from 
saturated olephins (alkanes)~\cite{fink-hans92jacs,smal+95jpc,holm+01jacs} to biological molecules
(such as DNA~\cite{kell-bart99sci,lewi+00jacs,fuku-tana98ac}).
Theoretically, 
the electronic mechanism underlying these experiments 
has been analyzed and understood.~\cite{holm+01jacs,tomf-sank02prb,nitz-ratn03sci,prod-car09prb}
As stated in Ref.~[\onlinecite{tomf-sank02prb}], 
the key parameter $\beta$ can be expressed ({\it eg} in MIM junctions) 
in terms of ($i$) the band gap $E_g$ and the (frontier) band widths
(or hopping parameter) $t$ of the insulating layer, 
and ($ii$) the alignment of the Fermi level in the metals with the
energy gap of the insulator.
Indeed, the effect of the electronic structure of the insulating layer
can be singled out by evaluating~\cite{tomf-sank02prb} the complex band structure (CBS), 
or evanescent states, in the limit of an infinitely long insulating region.
The CBS approach is also particularly interesting for an {\it ab initio} evaluation
of $\beta$, where the calculations can
be performed either using wavefunction-~\cite{pica+03jpcm} 
or Green's function-based~\cite{tomf-sank02prb} approaches.
A cartoon describing the relation between the electronic structure of the MIM junction 
and the evaluation of the $\beta$-decay factor is given in Fig.~\ref{fig:cartoon}.
Nevertheless, since the $\beta$ parameter can be~\cite{tomf-sank02prb}
directly related to the ratio between the energy gap and the band width of the insulator layer, the accuracy
of standard electronic-structure simulations based on the Kohn-Sham (KS) framework 
of the density functional theory (DFT) can be questioned.
In fact, using eigenvalues computed from KS-DFT it is well known that the fundamental band gap is badly underestimated
and (when using local and semilocal approximations) the delocalization of electronic states is typically overestimated. 
Moreover, the description of this class of experiments in terms of single particle
energies would require them to be interpreted as quasi-particle energies, in order to address the electronic
dynamics of the system. 
This is valid for advanced MBPT methods,~\cite{fett-wale71book} such as Hedin's 
GW approximation,~\cite{hedi65pr,hedi-lund69ssp}
and, at least in a perturbative sense, also for Hartree-Fock calculations.
However, DFT Kohn-Sham states are fictitious orbitals with
no direct physical interpretation, and their use in this context can
only be justified by the assumption that the exchange-correlation kernel is an
approximation for the quasi-particle Hamiltonian.
Model self-energies have also been recently~\cite{quek+09nl} 
used to correct the electronic structure of metal-molecule-metal junctions and found to be
important for the evaluation of $\beta$.
To date there has been no systematic investigation of the performance of different {\it ab initio} schemes in the
calculation of the $\beta$ decay factors.

\begin{figure}[!b]
   \includegraphics[clip, width=0.4\textwidth]{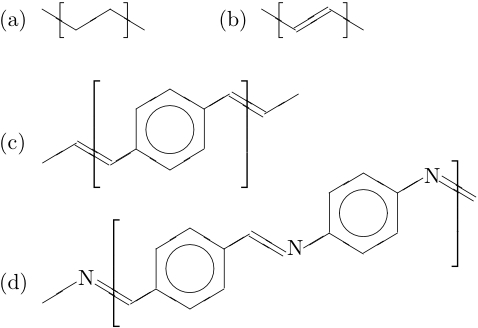}
   \caption{ \label{fig:polymers}
        Polymers studied:    (a) PE (poly-ethylene), with each Carbon atom in the polymer chain 
                             being fully saturated by two Hydrogen atoms; 
                             (b) PA (poly-acetylene), where sp$^2$ hybridization of Carbon atoms in the chain
                              (each one also bonded to one Hydrogen atom) implies $\pi$-conjugation, i.e. 
                              alternation of single and double C-C bonds along the chain;
                             (c) PPV [poly-({\it para} phenylene-vinylene)], constituted by benzene rings 
                               connected through vinyl groups, also presenting conjugation;
                             (d) PPI [poly-(phenylene-imine)], differing from PPV for the substitution of one Carbon atom 
                                in each vinyl group by a N atom, which still preserves the polymer conjugation. }
\end{figure}
In this paper we study
the effects of hybrid exchange-correlation functionals~\cite{kumm-kron08rmp} and 
the GW~\cite{hedi65pr,hedi-lund69ssp} approximation
on the calculation of the $\beta$ decay-factors, according to a scheme based on the
the complex band structure (CBS) formalism.~\cite{tomf-sank02prb}
A detailed comparison with local and semilocal functionals is also provided. 
The theoretical background of CBS, and GW and hybrid functionals
are described in Sec.~\ref{sec:method_transport} and ~\ref{sec:method_electronic_structure} respectively.
We discuss a general application of the Wannier functions interpolation to
the case of GW electronic structure (Sec.~\ref{sec:GW_interpolation}).
Our approach is applied to a number of polymers as reported
in Fig.~\ref{fig:polymers}: We compute the CBS for 
poly-ethylene (PE) and poly-acetylene (PA) as references for saturated and conjugated
chains respectively. We then consider poly-({\it para} phenylene-vinylene) (PPV)~\cite{burr+90nat}
and poly-(phenylene-imine)~\cite{choi+08sci,choi+10jacs} (PPI), which are relevant from a technological point of view, 
where we compare also with recent experimental 
data.~\cite{choi+08sci} 
All chains are studied as isolated, see App.~\ref{sec:details} for full numerical details.

\def\Xint#1{\mathchoice
{\XXint\displaystyle\textstyle{#1}}%
{\XXint\textstyle\scriptstyle{#1}}%
{\XXint\scriptstyle\scriptscriptstyle{#1}}%
{\XXint\scriptscriptstyle\scriptscriptstyle{#1}}%
\!\int}
\def\XXint#1#2#3{{\setbox0=\hbox{$#1{#2#3}{\int}$ }
\vcenter{\hbox{$#2#3$ }}\kern-.5\wd0}}
\def\ddashint{\Xint=}
\def\dashint{\Xint-}

\section{Method: Transport}
\label{sec:method_transport}

\label{sec:method_CBS}

To simulate the decay factors 
of non-resonant tunneling experiments
we adopt the 
CBS algorithm proposed by Tomfohr and Sankey (TS).~\cite{tomf-sank02prb}
For a recent discussion of the connection of transport properties with the CBS theory 
see also the work by Prodan and Car.~\cite{prod-car09prb}
Within the TS approach, we need to evaluate the CBS in the limit of an infinitely
thick
insulating region ($\beta$ is in fact an asymptotic behavior).
The 
outcome of this procedure is a set of $\beta(E)$ curves.
The value $\beta_0$ which has to be compared with the experiments 
is the smallest one ({\it i.e.} the most
penetrating) aligned with the Fermi level of the junction:
\begin{equation}
   \beta_0  = \beta(E_F)
\end{equation}
Since the electrodes are not considered in the calculation, together with the proper metal-insulator
interface, $E_F$ is not known a priori and must be either estimated or calculated separately.
This issue is discussed in detail in Ref.~[\onlinecite{tomf-sank02prb}]. In the present work we will
not compute the Fermi level alignment explicitly, and will rely on the estimation method proposed in the
above reference,~\cite{tomf-sank02prb} 
which consists in evaluating $\beta(E)$ at the energy where $d\beta/dE = 0$ (branch point). 
This method is originally due to 
Tersoff~\cite{ters84prl} and based on the pinning of $E_F$ by metal induced gap states (MIGS).
We discuss this approximation in Sec.~\ref{sec:discussion} where we compare
with experimental results.
Note that the value of $\beta$ at the branch point is also connected (for non-metallic 
1D systems with a local potential) 
with the degree of localization of the density matrix, since $\beta$ determines~\cite{kohn59pr,he-vand01prl}
its spatial decay. Pictorially, a better description of $\beta(E)$ means then
an improved description of the electronic localization.

Scanning the energy spectrum, the CBS procedure searches for 
evanescent solutions to a given effective single-particle Hamiltonian.
By definition, states with (real) Bloch symmetry $\mathbf{k}$ 
satisfy the relation 
\begin{equation}
   \label{eq:translation}
   \hat{T}(\mathbf{R}) \, \psi_{\mathbf{k}}(\mathbf{r}) = \psi_{\mathbf{k}}(\mathbf{r}+\mathbf{R}) = 
   \lambda \, \psi_{\mathbf{k}}(\mathbf{r}), 
\end{equation}
where 
$\mathbf{R}$ is any direct lattice vector, $\hat{T}(\mathbf{R})$ a translation operator, and
$\lambda = e^{i\mathbf{k} \cdot \mathbf{R}}$.
In the same way it is possible to define a {\it complex} Bloch symmetry 
$\boldsymbol{\kappa}=\mathbf{k}+i\boldsymbol{\beta}/2$
setting 
$\lambda = e^{i\boldsymbol{\kappa} \cdot \mathbf{R}} = e^{i\boldsymbol{k} \cdot \mathbf{R}} \, 
e^{-\boldsymbol{\beta} \cdot \mathbf{R}/2} $, which is thus no longer a pure phase.
The imaginary part of $\boldsymbol{\kappa}$ implies
a real space exponential decay of the wavefunctions and it is
customary to define $\beta(E) = 2\left|\text{Im}[\boldsymbol{\kappa}(E)\cdot \hat{e}]\right| $ 
($\hat{e}$ transport direction).
The energy dependence of $\kappa$ comes from the fact that
for a fixed energy $E$, the solutions are searched in terms of $\boldsymbol{\kappa}$, as it is
usually done in scattering theory.

By adopting a localized basis set $\{ | \phi_{i\mathbf{R}} \rangle \}$ ($i$ and $\mathbf{R}$ orbital and 
lattice indexes, respectively), 
it is possible to define
the Hamiltonian ${H}$ and overlap ${S}$ operators through their matrix elements $H_{ij}(\mathbf{R}) = 
\langle \phi_{i\mathbf{0}} | {H} | \phi_{j\mathbf{R}} \rangle $ and
$S_{ij}(\mathbf{R}) = \langle \phi_{i\mathbf{0}} | \phi_{j\mathbf{R}} \rangle $, and the 
wavefunctions as $ | \psi_l \rangle = \sum_{i\mathbf{R}} C_{il}(\mathbf{R})\, |\phi_{i\mathbf{R}}\rangle$.
Setting ${Z} = {H} -E\,{S} $, the eigenvalue equation for ${H}$ can be written as:
\begin{equation}
   \sum_{m=-N}^{N} \, Z(\mathbf{R}_m)  C(\mathbf{R}_m) = 0,
\end{equation}
where matrix multiplication is implicit, and $N$ is the number defining the last non-zero
matrix $Z(\mathbf{R}_N)$. Here we are assuming a real space decay of the Hamiltonian and
overlap matrices, which is typically physical even if the range can be strongly dependent on the
scheme used to define the Hamiltonian. We will comment later on this point when discussing the 
use of GW or HF methods.
It is then possible to derive~\cite{tomf-sank02prb} 
the following system of $2N$ matrix equations:
\begin{eqnarray}
   \label{eq:cbs_matrix}
   -\sum_{m=-N}^{N-1} \, Z(\mathbf{R}_m)  C(\mathbf{R}_m) =  Z(\mathbf{R}_N) \,  \lambda C(\mathbf{R}_{N-1}), \\
    C(\mathbf{R}_{m+1}) = \lambda C(\mathbf{R}_{m}), \qquad m=-N,N-2.
   \label{eq:cbs_matrix2}
\end{eqnarray}
where Eq.~(\ref{eq:cbs_matrix2}) is a direct consequence of Eq.~(\ref{eq:translation}).
Such matrix equations become an eigenvalue problem for $\lambda$ assuming we can invert the matrix
$Z(\mathbf{R}_N)$. As pointed out in Ref.~[\onlinecite{tomf-sank02prb}], this is very often not the case as
the matrix is singular,
but the singluarities can be avoided.
The reader is referred to the original work for the details.
The full algorithm proposed in Ref.~[\onlinecite{tomf-sank02prb}] has been implemented in
the \textsc{WanT} code~\cite{WanT,ferr+07jpcm} and used for the present work.
A simple tight-binding analytical model (a generalized version of the one presented in
Ref.~[\onlinecite{tomf-sank02prb}]) is discussed in App.~\ref{sec:model}.
This model will be used in Sec.~\ref{sec:results}
to fit and interpret the real and complex band-structures of the polymers we have investigated here.

\section{Method: Electronic structure}
\label{sec:method_electronic_structure}

According to the above discussion, in order to simulate the decay coefficient of a MIM junction, 
we need to compute the electronic structure of the insulating layer (considered as infinitely extended).
The underlying reason for this simplification is that we interpret the computed single-particle energies
of the system as quasi-particle (QP) energies, which in turn determine the dynamics of singly charged
excitations. In general the transport problem for interacting systems 
is more complicated than that and requires a more sophisticated treatment.~\cite{sai+05prl,ferr+05prl,ferr+05prb,boke+07prb,%
kurt+05prb,thyg-rubi08prb,vign-dive09prb,ness+10prb,kurt+10prl}
For DFT, the understanding of how the exact KS-DFT Hamiltonian performs to compute transport 
has been recently the subject of several investigations.~\cite{tohe+05prl,ke+07jcp,mera+10prb,mera-niqu10prl}
Apart from the properties of the exact functional, 
currently available DFT approximations like LDA or GGA have been demonstrated
to systematically overestimate the conductance, especially for off-resonant 
junctions.~\cite{sai+05prl,lind-ratn07amat,dive09amat}
Pragmatically, this suggests that corrections beyond local and semilocal KS-DFT approaches are needed.

Using QP-corrected electronic structure to compute charge transport ({\it eg} by means of
the Landauer formula~\cite{land70pm}) through interacting systems
seems to be a reasonable
approximation when finite-lifetime effects are weak.~\cite{datt-tian97prb,ness+10prb}
Indeed, a number of works computing QP energies by means of the 
GW approximation~\cite{dara+07prb,thyg-rubi07jcp,thyg-rubi08prb,stra+11prb,rang+11prb}
or model self-energies~\cite{quek+07nl,ceho+08prb,mowb+08jcp,quek+09nl}
have been reported in the literature.
On the other hand,
hybrid-exchange functional methods like B3LYP~\cite{step+94jpc} or PBE0~\cite{perd+96jcp}
are widely used and lead to band gaps and band widths which are
usually closer~\cite{musc+01cpl} to the experimental values than simple semi-local KS approaches, for both
molecules and solids. Recent works~\cite{rost+10prb,jain+11prl,refa+11prb,blas+11prb} have further investigated 
the accuracy of hybrid exchange functionals, also in comparison with GW calculations.
In this work we compare GW and hybrid functionals 
for the calculation of electronic structure and transport properties of selected organic polymers.
In the following we summarize the GW approximation and underline some formal similarities with
hybrid functionals.

\subsection{The GW approximation and hybrid DFT}

A many-body theoretical formulation of the electronic structure problem can be
obtained by using the Green's function formalism.
The one-particle excitation energies of an interacting system are
the poles of its interacting Green's function ${G}(E)$,~\cite{fett-wale71book,onid+02rmp} 
which can be written as:
\begin{equation}
   {G}(E) = \left[ E {I} -{h}_0 -{\Sigma}(E) \right]^{-1},
\end{equation}
where ${h}_0$ is an effective single particle Hamiltonian and
${\Sigma}(E)$ is the non-local, non-hermitean, frequency dependent self-energy operator.
In general, $\Sigma$ is not known {\it a priori} and must be approximated. In this work
the self-energy is computed within the GW 
approximation:~\cite{hedi65pr,hedi-lund69ssp}
\begin{eqnarray}
   \nonumber
   \Sigma_{\text{GW}}(\mathbf{r}_1,\mathbf{r}_2,E) = i \int \frac{d\omega'}{2\pi} \, e^{-i\delta\omega'} 
                                          G(\mathbf{r}_1,\mathbf{r}_2,E-\omega') \, \times \\
           W(\mathbf{r}_1,\mathbf{r}_2,\omega'),
   \label{eq:sigma_GW}
\end{eqnarray}
where ${W}(\omega)$ is the screened Coulomb interaction evaluated in the random phase approximation (RPA). 
For more details see e.g. Refs.~[\onlinecite{hybe-loui86prb1,onid+02rmp}].
In the simplest implementation of the GW approximation, the self-energy is computed non self-consistently, {\it i.e.} 
by evaluating ${G}$ and ${W}$
according to the eigenvalues and eigenvectors of a reference non-interacting Hamiltonian 
(typically the KS Hamilotonian at the LDA or GGA level).
Such a procedure is known as $G_0W_0$ and gives reasonable results
for the quasiparticle energies in a number of cases.~\cite{hybe-loui86prb,arya-gunn98rpp,onid+02rmp} 
In the present paper we exploit the $G_0W_0$ approximation, and evaluate the frequency integrals in 
Eq.~(\ref{eq:sigma_GW}) by using a plasmon pole model according to Godby and Needs.~\cite{godb-need89prl}

In this work, the main quantities we are interested in are the QP energies.
If we neglect finite lifetime effects and take (or symmetrize) the self-energy to be hermitean,
QP energies are given by first order perturbation theory as:
\begin{eqnarray}
   \label{eq:qp_energy}
   \epsilon^{\text{QP}}_m &=& \epsilon^{\text{KS}}_m + 
            \langle \psi_m | \, {\Sigma}(\epsilon^{\text{QP}}_m) -{v}_{\text{xc}} \, | \psi_m \rangle, 
\end{eqnarray}
As customary,~\cite{hybe-loui86prb1} 
in order to solve for $\epsilon^{\text{QP}}_m$, 
the self-energy in the above equation is expanded to first order as a function of $E$.

In order to get more physical insight
we also refer to the (static) COHSEX~\cite{hedi65pr,hybe-loui86prb1} approximation of GW, where 
the self-energy is written as $\Sigma = \Sigma^{\text{COH}} + \Sigma^{\text{SEX}}$:
\begin{eqnarray}
     \label{eq:sigma_SEX}
     \Sigma^{\text{SEX}}(\mathbf{r}_1,\mathbf{r}_2) &=& -\gamma(\mathbf{r}_1,\mathbf{r}_2) \, 
                W(\mathbf{r}_1,\mathbf{r}_2,0), \\
     \label{eq:sigma_COH}
     \Sigma^{\text{COH}}(\mathbf{r}_1,\mathbf{r}_2) &=& \frac{1}{2} \, \delta(\mathbf{r}_1,\mathbf{r}_2)
                \, W_p(\mathbf{r}_1,\mathbf{r}_2,0).
\end{eqnarray}
Here $\gamma(\mathbf{r}_1,\mathbf{r}_2)$ is the one particle density matrix, and $W_p = W -v$ is the dynamical
contribution to $W$ (being $v$ the bare Coulomb interaction).
The COHSEX self-energy is thus the sum of a statically screened exchange term and a local potential.
This partition is particularly useful for discussing the connection between hybrid exchange 
functionals and GW. In the former case, the potential can be written as:~\cite{kumm-kron08rmp}
\begin{equation}
   \label{eq:hybrids_energy}
   v^{\text{hyb}}_{\text{xc}} = \alpha \, v^{\text{NL}}_{\text{x}}
                                + (1-\alpha) \, v^{\text{L}}_{\text{x}}
                                + v^{\text{L}}_{\text{c}},
\end{equation}
where $v^{\text{NL}}_{\text{x}}$ is the non-local exchange potential, while $v^{\text{L}}_{\text{x}}$ and
$v^{\text{L}}_{\text{c}}$ are local potentials. It is then straightforward to 
interpret $\alpha$ as an inverse effective screening to stress the formal analogy of
Eqs.~(\ref{eq:sigma_SEX}--\ref{eq:sigma_COH})
and (\ref{eq:hybrids_energy}). Similar considerations are of course valid for more complex forms of hybrid 
functionals, like range separated or local formulations.~\cite{kumm-kron08rmp,heyd+03jcp,perd+07pra}
This formal analogy is well-known in the literature,~\cite{kumm-kron08rmp} and it has also been further investigated 
recently.~\cite{marq+11prb} 
Besides their accuracy for thermochemistry, this analysis highlights that also the electronic structure computed
by non-local hybrids can benefit from the inclusion of some screened exchange term. Indeed, improved 
description of the electronic
structure for finite and extended systems are typically found,~\cite{musc+01cpl,marq+11prb} 
even though the accuracy may vary significantly depending on the system.
\subsection{Interpolation of GW electronic structure using Wannier functions}
\label{sec:GW_interpolation}

Dealing with periodic systems, interpolation over 
the first Brillouin zone (1BZ) is a long standing issue. 
Calculations are typically performed by discretizing $\mathbf{k}$-points 
in the 1BZ, and some post-processing schemes
[such as {\it eg} those to compute density-of-states (DOS), Fermi surface, band 
structure, or phonons] might need a better discretization of 1BZ than some of the previous steps
(often aimed at computing total energy, forces, and charge density).
This is particularly critical when the computational requirements of the adopted methods limit the 
$\mathbf{k}$-point discretization, as is the case for GW calculations.
Schemes able to refine or interpolate~\cite{bloc+94prb} over the 1BZ are particularly useful
for this purpose.
One of these is the {\it Wannier interpolation},~\cite{marz-vand97prb,souz+01prb,yate+07prb,ferr+07jpcm}
where the localization of the Wannier function (WF) basis together with the 
finite range in real space of the Hamiltonian
are used to perform a Fourier interpolation of the eigenvalues, and eventually eigenvectors.
The use of this scheme to interpolate GW results has been also reported elsewhere by 
Hamann and Vanderbilt.~\cite{hama-vand09prb}

The procedure can be applied not only to the Hamiltonian, but in principle to any operator $A(\mathbf{r},\mathbf{r}')$
with the translational symmetry of the Hamiltonian.
First, we define the projector $P$ over a subspace of interest:
\begin{equation}
   P = \sum_{n\mathbf{k}} \, | \psi_{n\mathbf{k}} \rangle \langle \psi_{n\mathbf{k}} |, 
\end{equation}
in terms of the eigenvectors of $H$.
When the subspace $P$ is complete, we can represent $A$ as:
\begin{eqnarray}
   \label{eq:WF_repres1}
   A &=& \sum_{\mathbf{k}} \sum_{mn} | \psi_{m\mathbf{k}} \rangle \, A_{mn}(\mathbf{k}) \, \langle \psi_{n\mathbf{k}} | \\
   \label{eq:WF_repres2}
   A_{mn}(\mathbf{k}) &=& \langle \psi_{m\mathbf{k}} | A | \psi_{n\mathbf{k}} \rangle
\end{eqnarray}
Here, $A$ is diagonal with respect to the $\mathbf{k}$-index because it commutes with the 
translation operators of the direct lattice (as assumed).
In practice, limiting the number of eigenstates of $H$ included in $P$
is equivalent to considering the projection of $A$ on the $P$ subspace, namely $A^P = P A P$ instead of $A$. 
At this point we can use the definition of maximally localized Wannier functions (MLWF's), 
according to Ref.~[\onlinecite{marz-vand97prb}]:
\begin{equation}
  | w_{i\mathbf{R}} \rangle = \frac{1}{N_{\mathbf{k}}} \sum_{\mathbf{k}} \, e^{-i\mathbf{k}\mathbf{R}} \, \sum_m \, 
                     U^{\mathbf{k}}_{mi} \, | \psi_{m\mathbf{k}} \rangle,
\end{equation}
to obtain an expression for the matrix elements of $A$ on the Wannier 
basis, $A^P_{ij}(\mathbf{R}) =  \langle w_{i\mathbf{0}} | A | w_{j\mathbf{R}} \rangle $:
\begin{equation}
    \label{eq:WFs_k2R}
    A^P_{ij}(\mathbf{R}) =  
                          \frac{1}{N_{\mathbf{k}}} \sum_{\mathbf{k}} \, e^{-i\mathbf{k}\mathbf{R}} \,
                         \left[ U^{\mathbf{k}\dagger} A^P(\mathbf{k}) U^{\mathbf{k}} \right]_{ij} .
\end{equation}
Note that when the original Marzari-Vanderbilt procedure~\cite{marz-vand97prb} 
is applied without any disentanglement,~\cite{souz+01prb} the $U^{\mathbf{k}}$ matrices are a unitary mapping
of $N$ Bloch states (usually occupied, but not necessarily) into $N$ WF's. 
Instead, when the disentanglement is performed, 
the resulting WF set does not span the whole
$P$ subspace ($U^{\mathbf{k}}$ are then rectangular matrices). This means that in the general case 
the final representation of $A$ is actually 
not projected on $P$ but on the smaller subspace spanned by the WFs.

Assuming that $A^P$ is decaying fast enough in real space to have $||A^P(\mathbf{R})||\simeq 0$ for
$|\mathbf{R}| > |\mathbf{R}_0|$ (where $\mathbf{R}_0$ is within the finite set compatible with the
initial $\mathbf{k}$-point grid), we can perform the following Fourier interpolation to obtain the
matrix elements of $A$ for any $\mathbf{k}'$ point:
\begin{equation}
   A^P_{ij}(\mathbf{k}') = \sum_{\mathbf{R}}^{|\mathbf{R}|<|\mathbf{R_0}|} 
                          \, e^{i\mathbf{k}'\mathbf{R}} \, A^P_{ij}(\mathbf{R})
\end{equation}

When interpolating GW results, we want to represent the operator $\Sigma(E) = 
\Sigma^{GW}(E) -v_{\text{xc}}$
which is in general non-local, non-hermitean and frequency dependent. For the sake of the Wannier interpolation,
we are mainly interested to check that the intrinsic non-locality of $P \Sigma P$ is compatible with the selected
$\mathbf{k}$-point grid (or, in other terms, that the GW calculation is converged with respect to the number of
$\mathbf{k}$-points used).
The localization of the GW self-energy is further discussed
in Sec.~\ref{sec:GW_data}, especially in connection with the usual approximation that neglects
off-diagonal $\Sigma_{mn}(\mathbf{k}E)$ matrix elements.

\subsection{Numerical approach}

\label{sec:numerical_approach}
In this work, DFT and hybrid-DFT calculations have been performed using the {\sc CRYSTAL09} 
package.~\cite{CRYSTAL09} The code implements all-electron electronic structure methods within 
periodic boundary conditions and adopts
an atomic basis set expanded in Gaussian functions (further details in App.~\ref{sec:details}).
Once the Hamiltonian matrix elements are obtained,~\cite{note_ortho_crystal} 
the real and complex band structures are interpolated (as discussed in the previous Sections) 
using the \textsc{WanT}~\cite{WanT,ferr+07jpcm} package.  

GW results have been obtained using the plane-wave and pseudopotentials implementation
of {\sc SaX}~\cite{mart-buss09cpc} which is interfaced to {\sc Quantum-ESPRESSO}~\cite{gian+09jpcm} (QE) 
for DFT calculations.
In this case, once the Kohn-Sham electronic structure is evaluated,  
we first compute MLWF's~\cite{marz-vand97prb,souz+01prb} using 
{\sc WanT} and then apply the CBS technique.
In order to assess any systematic error in comparing GW and hybrid-DFT results (which have been
obtained using different basis sets such as plane waves and local orbitals), 
we have also performed hybrid-DFT and HF calculations using QE and {\sc SaX}. In this case WFs are computed
on top of the already corrected electronic structure.
Results are shown for the
case of polyacetylene (see Tab.~\ref{tab:betas_pa} and Fig.~\ref{fig:GW_offdiag}(b) in particular).
The excellent agreement between the two sets of data suggests that the pseudopotential approximation, the 
basis set, and the numerical thresholds are sufficiently well converged to have negligible influence
on the results presented.
Full computational details and parameters are reported in App~\ref{sec:details}.
\section{Results}
\label{sec:results}

\subsection{Structural properties}
\label{sec:structure}

Before focussing on the electronic and transport properties of the isolated polymer chains from Fig.~\ref{fig:polymers}, 
we investigate
their structure by fully relaxing both the atomic positions and the cell parameters using 
different exchange-correlation schemes. 
All systems are treated with one-dimensional periodicity, the details of the calculations 
(performed using {\sc CRYSTAL09}) have been given in App.~\ref{sec:details}.
The results for the lattice parameters are reported in Tab.~\ref{tab:geometries}.
\begin{table}
   \centering
   \caption{\label{tab:geometries}
            Lattice parameter $c$ [\AA] for PA, PE, PPV, and PPI, computed using different
            XC schemes as implemented in {\sc CRYSTAL09}. In the case of PA, we also report
            the bond length alternation [\AA] (BLA) of single and double C-C bonds.
   }
   \begin{ruledtabular} 
   \begin{tabular}{ l | cc | ccc }
   \vspace{2pt}
   Scheme &  {\bf PA} & PA-BLA  & {\bf PE} &   {\bf PPV} &  {\bf PPI} \\
   \hline    
   \hline    
   \vspace{2pt}
   LDA      &   2.463   & 1.369/1.416 &  2.537   &  6.644   &  12.869   \\[5pt]
   PW       &   2.482   & 1.376/1.428 &  2.570   &  6.712   &  13.001   \\
   BLYP     &   2.493   & 1.380/1.435 &  2.590   &  6.747   &  13.079   \\
   PBE      &   2.484   & 1.378/1.429 &  2.572   &  6.719   &  13.014   \\[5pt]
   B3PW     &   2.471   & 1.362/1.432 &  2.560   &  6.692   &  12.961   \\
   B3LYP    &   2.476   & 1.375/1.435 &  2.570   &  6.706   &  12.998   \\
   PBE0     &   2.468   & 1.359/1.432 &  2.553   &  6.680   &  12.938   \\[5pt]
   HF       &   2.465   & 1.332/1.457 &  2.556   &  6.689   &  12.950   \\[5pt]

   \end{tabular}
   \end{ruledtabular}
\end{table}

In the case of PA, electronic properties such as the band gap (as well as the 
evanescent states) are strongly dependent on the dimerization of the C-C bond lengths
(Peierls distortion).
Such bond alternation is not easily captured by local and semilocal DFT schemes leading
to band gaps that are far too small, i.e. to an overestimation of the metallicity.
Since these parameters are critical to our purpose, we report also the bond length alternation (BLA)
for PA in Tab.~\ref{tab:geometries}. Our HF results are in excellent agreement with previously published 
results.~\cite{vanf+02prl,kirt+95jcp} In the following we will label as PA$_{\text{HF}}$ the calculations
performed 
using the geometry from Ref.~[\onlinecite{vanf+02prl}], where $c=2.469$\AA{} and BLA is 1.339/1.451 \AA.
We also decided to consider two frozen geometries of PA (namely PA$_1$ and PA$_2$) according to
other data in the literature.~\cite{rodr-lars95jcp,tomf-sank02prb}
In the case of PA$_1$\cite{rodr-lars95jcp} 
we set $c=2.451$ \AA{} and BLA to 1.370/1.460 \AA, while for PA$_2$\cite{tomf-sank02prb} $c=2.496$ \AA{}
and BLA to 1.340/1.540 \AA.
In passing we note that the PA$_1$ geometry is also very similar to the one adopted in 
Refs.~[\onlinecite{rohl-loui99prl,rohl+01sm,tiag+04prb}], where $c=2.451$ \AA{} and BLA is set to 1.360/1.440 \AA,
according to experimental data.~\cite{yann-clar83prl,zhu+92ssc} Since these theoretical studies report GW results, 
in Sec.~\ref{sec:electronic_structure_prototype} we will compare with PA$_1$.

Data from multiple geometries of PA are useful to decouple 
the electronic and structural effects 
of the adopted XC schemes on the complex band structure.
Since the extent of this goes beyond PA, in the
following we decided to look at the effect of XC treatment first using a frozen geometry,
independently of the adopted method, and then to compare also with the same
quantities obtained using fully relaxed geometries.
Moreover, since GW corrections are usually computed without performing a further
structural relaxation, working at fixed geometry allows us to compare GW corrections 
and results from hybrid functionals 
for identical geometries.
For each polymer except PA we adopted the geometry obtained after full relaxation
at the PBE level using {\sc Quantum-ESPRESSO}. Lattice parameters (2.564 \AA{} for PE, 6.702 \AA{} for PPV
and 13.004 \AA{} for PPI)
are in very good agreement with those in Tab.~\ref{tab:geometries} obtained using PBE in {\sc CRYSTAL09}.

\subsection{Electronic structure: poly-ethylene and poly-acetylene}
\label{sec:electronic_structure_prototype}

\begin{figure}[!t]
   \includegraphics[clip, width=0.45\textwidth]{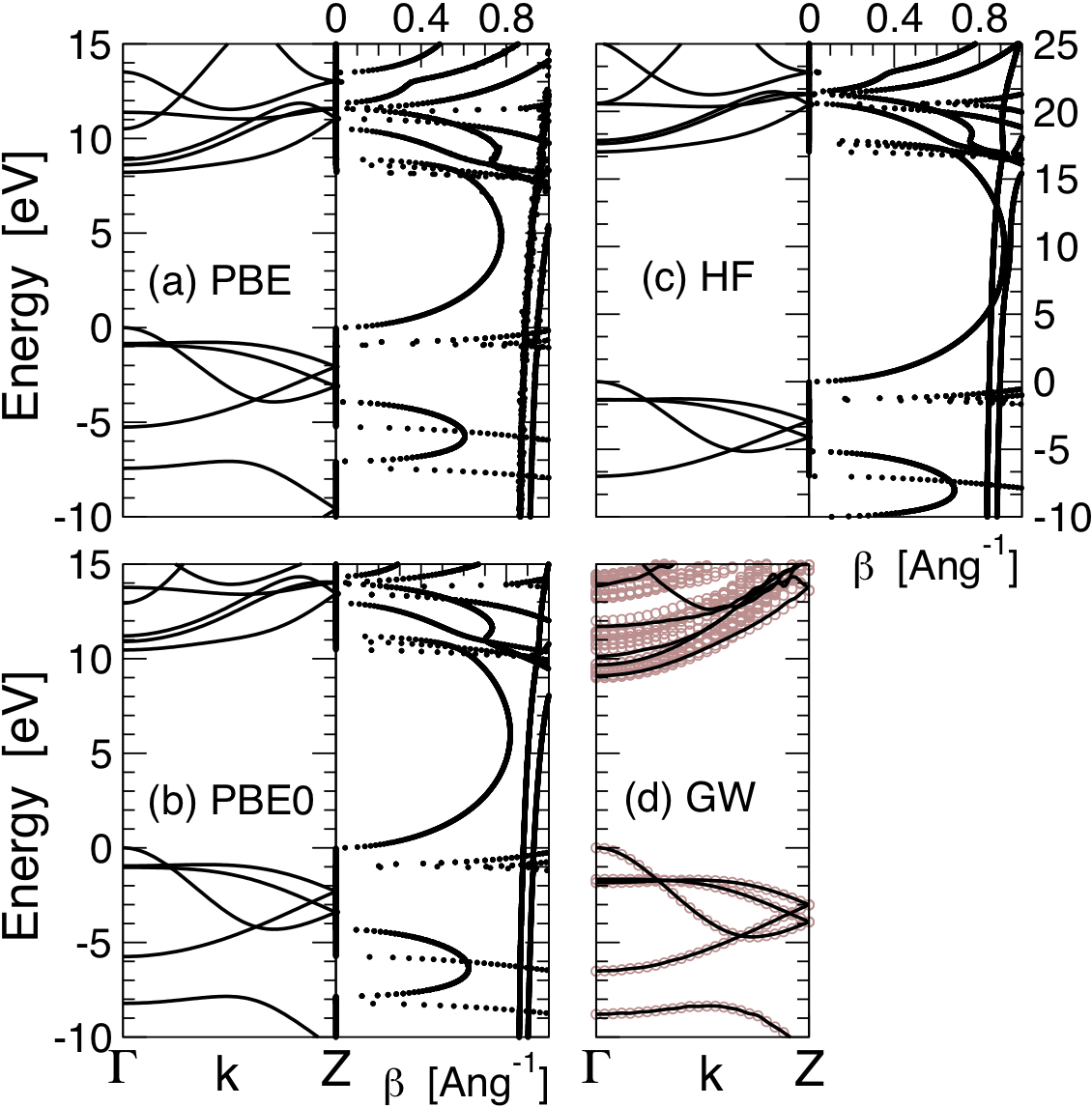}
   \caption{\label{fig:PE_bands}%
   Poly-ethylene (PE). Real and complex band structure using the 
   schemes: (a) PBE, (b) PBE0, (c) HF and (d) GW.
   Solid (black) lines in (a,b,c) refer to {\sc CRYSTAL09} calculations.
   GW results are obtained using {\sc SaX} (circles), and interpolated with WFs
   (solid black lines).
   }
\end{figure}

\begin{figure}
   \includegraphics[clip, width=0.45\textwidth]{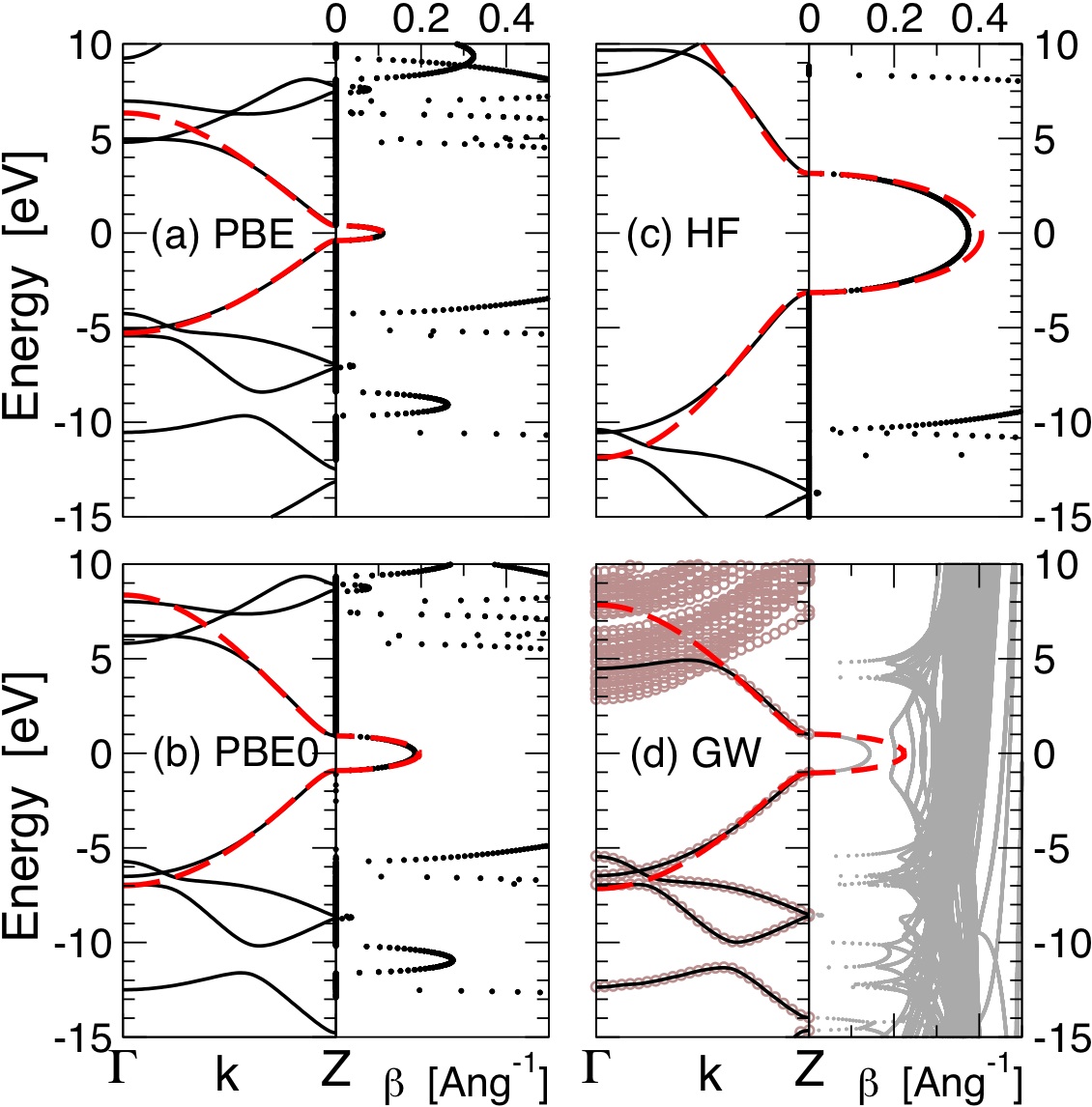}
   \caption{\label{fig:PA1_bands} (color online). Poly-acetylene (geometry PA$_1$).
   Real and complex band structure using the 
   schemes: (a) PBE, (b) PBE0, (c) HF and (d) GW.
   Solid (black) lines in (a,b,c) refer to {\sc CRYSTAL09} calculations.
   GW results are obtained using {\sc SaX} (circles), and interpolated with WFs (solid black lines).
   Dashed (red) lines refer to data interpolated according to 
   the tight-binding model in App.~\ref{sec:model}.}
\end{figure}

In this section we investigate the effect of different XC schemes (local and semilocal DFT, 
hybrid functionals, HF, and GW) on the electronic properties of two prototype polymers (PE and PA), 
while results for PPV and PPI will be reported in Sec.~\ref{sec:electronic_structure_applications}.
We compute both the real and the complex band structures. Figures~\ref{fig:PE_bands}-\ref{fig:PA1_bands} 
refer to the case of frozen geometries (i.e. geometry is not changing according to the adopted scheme, see also 
Sec.~\ref{sec:structure}). 
Details about the electronic structure of the three PA geometries studied are reported in Tab.~\ref{tab:betas_pa}.
In the case of PA (as well as PPV), we can also compare with previously published 
GW calculations.~\cite{rohl-loui99prl,rohl+01sm,pusc-ambr02prl,tiag+04prb}
Note that the GW results are interpolated using MLWF's,
which results in filtering out some of the states above the vacuum level.

In Fig.~\ref{fig:PE_bands}, we report~\cite{note_ortho_crystal} the real and complex band structure for PE.
Our results are in reasonably good agreement with previously 
published theoretical data.~\cite{tomf-sank02prb,pica+03jpcm,prod-car08nl,prod-car09prb,stra-thyg11bjn}
As can be seen from panels (a,b), the maximum value of $\beta$ from the arc across the fundamental gap
is not changing much when passing from PBE to PBE0 ($\beta_{\text{max}}\simeq0.8$ \AA$^{-1}$). %
In agreement with Ref.~[\onlinecite{prod-car09prb}], we believe this to be one of the reasons why
the $\beta$ computed at the LDA/GGA levels have been found in agreement with the experiment.

\begin{table}[!b]
   \centering
   \caption{\label{tab:betas_pe}
             PE: maximum of $\beta(E)$ [\AA$^{-1}$] inside the gap computed by means of
             different theoretical schemes. Results are obtained by \textsc{CRYSTAL09} (CRY)
             and by {\sc Quantum-ESPRESSO} (QE)--{\sc SaX}.
   }
   \begin{ruledtabular} 
   \begin{tabular}{ r | ccccc }
     $\beta_{\text{max}}$   &  \multicolumn{5}{c}{ poly ethylene (PE) } 
          \\[2pt]
   \hline    
   & & & & & \\[-7pt]
            &  LDA    &  PBE   &  PBE0  &   HF  &  COHSEX  \\[2pt]  %
   \hline    
   \hline
   \vspace{2pt}
   & & & & & \\[-7pt]                                     %
   CRY      &  0.77   &  0.77  &  0.81  &  0.92 &          \\[6pt] %
   QE-SaX   &  0.80   &  0.81  &        &  0.94 &   0.94   \\     
   \end{tabular}
   \end{ruledtabular}
\end{table}

The G$_0$W$_0$-corrected band structure is reported in panel (d). It compares favorably with existing 
literature data.\cite{pica+03jpcm} 
The main qualitative difference with Fig.~2(a) of Ref.~[\onlinecite{pica+03jpcm}] comes from the filtering of
some vacuum-like states, due to the use of the localized basis (or WF interpolation) 
in our approach. Our results are more similar to those presented in Ref.~[\onlinecite{tomf-sank02prb}],
obtained using a localized basis set. This has little influence on
the part of the CBS spectrum that is physically relevant for the non-resonant tunneling experiment.
The G$_0$W$_0$ CBS is not computed here because the missing self-consistency is critical for CBS, 
as it is discussed in Sec.~\ref{sec:GW_data}. 
Instead, we have computed CBS from the self-consistent~\cite{note_COHSEX} COHSEX
electronic structure. %
All results for $\beta_{\text{max}}$ are collected in Tab.~\ref{tab:betas_pe}, where we compare data
obtained by using \textsc{CRYSTAL} and \textsc{Quantum-ESPRESSO} or \textsc{SaX} after Wannierization. 
Results and trends compare reasonably well.

\begin{table}
   \centering
   \caption{\label{tab:betas_pa}
            PA$_{\text{HF}}$, PA$_1$, PA$_2$: Band gap ($E_g$, [eV]), hopping parameters
            ($t_1$, $t_2$, [eV]) according to tight-binding (TB) model in Eq.~(\ref{eq:model_3p}), 
            maximum of $\beta(E)$ [\AA$^{-1}$] inside the gap.
            Calculations are performed using DFT, hybrid-DFT, HF, and diagonal-GW schemes. 
            $\beta_{\text{max}}$ through the CBS interpolation by using
            the TB model are also shown.
            All the data have been computed using {\sc CRYSTAL09} except the
            GW results and the X-QE lines (X=LDA,PBE,PBE0).
   }
   \begin{ruledtabular} 
   \begin{tabular}{ l | ccc | cc }
   \vspace{2pt}
   Scheme &  $\mathbf{E_g}$ & $\mathbf{t_1}$ & $\mathbf{t_2}$ &
             $\boldsymbol{\beta_{\text{max}}}$ & $\boldsymbol{\beta^{\text{model}}_{\text{max}}}$ \\[2pt]
   \hline
   \hline    
       \multicolumn{6}{c}{ poly-acetylene (PA$_{\text{HF}}$) } \\[2pt]
   \hline
   \vspace{2pt}
   LDA      &   0.99     &  2.97   &  0.199   &   0.13   &  0.13  \\[5pt]
   PW       &   1.01     &  2.97   &  0.203   &   0.14   &  0.14  \\
   BLYP     &   1.02     &  2.96   &  0.199   &   0.14   &  0.14  \\
   PBE      &   1.01     &  2.98   &  0.204   &   0.14   &  0.14  \\[5pt]
   B3PW     &   1.94     &  3.78   &  0.223   &   0.19   &  0.21  \\
   B3LYP    &   1.95     &  3.77   &  0.218   &   0.20   &  0.21  \\
   PBE0     &   2.20     &  3.98   &  0.228   &   0.21   &  0.22  \\[5pt]
   HF       &   6.97     &  6.45   &  0.277   &   0.38   &  0.43  \\[5pt]
   d-G$_0$W$_0$ &  2.60    &  3.85   &   0.154  &   (0.17)  &  0.27 \\[5pt]
   \hline
   \vspace{2pt}
   LDA-QE   &   0.99     &  3.02   &  0.214   &   0.13   &  0.13  \\
   PBE-QE   &   1.02     &  3.00   &  0.212   &   0.14   &  0.14  \\
   PBE0-QE  &   2.22     &  3.99   &  0.238   &   0.20   &  0.22  \\
   \hline
   \hline
       \multicolumn{6}{c}{ poly-acetylene (PA$_1$) } \\[2pt]
   \hline
   \vspace{2pt}
   LDA      &   0.78     &  2.90   &  0.144   &  0.11   &  0.11  \\[5pt]
   PW       &   0.80     &  2.90   &  0.148   &  0.11   &  0.11  \\
   BLYP     &   0.80     &  2.89   &  0.144   &  0.11   &  0.11  \\
   PBE      &   0.80     &  2.91   &  0.149   &  0.11   &  0.11  \\[5pt]
   B3PW     &   1.64     &  3.73   &  0.166   &  0.17   &  0.18  \\
   B3LYP    &   1.64     &  3.72   &  0.161   &  0.17   &  0.18  \\
   PBE0     &   1.88     &  3.94   &  0.171   &  0.18   &  0.19  \\[5pt]
   HF       &   6.46     &  6.46   &  0.212   &  0.36   &  0.40  \\[5pt]
  d-G$_0$W$_0$ &  2.05     &  3.72   &  0.090   &  (0.14)   &  0.22  \\
   \hline
   \hline    
       \multicolumn{6}{c}{ poly-acetylene (PA$_2$) } \\[2pt]
   \hline
   \vspace{2pt}
   LDA      &   1.68     &  2.76   &  0.134  &  0.24   &  0.24   \\[5pt]
   PW       &   1.72     &  2.76   &  0.139  &  0.25   &  0.25   \\
   BLYP     &   1.72     &  2.75   &  0.134  &  0.25   &  0.25   \\
   PBE      &   1.72     &  2.77   &  0.139  &  0.25   &  0.25   \\[5pt]
   B3PW     &   2.89     &  3.49   &  0.153  &  0.31   &  0.33   \\
   B3LYP    &   2.90     &  3.47   &  0.149  &  0.31   &  0.33   \\
   PBE0     &   3.20     &  3.67   &  0.158  &  0.33   &  0.35   \\[5pt]
   HF       &   8.37     &  5.88   &  0.196  &  0.50   &  0.56   \\[5pt]
  d-G$_0$W$_0$ &  4.17     &  3.87   &  0.092   &  (0.27)   &  0.43  \\
   \end{tabular}
   \end{ruledtabular}
\end{table}

In Fig.~\ref{fig:PA1_bands}, 
we show the results for real and complex band structure in the case
of PA$_1$. 
The dependence of the $\beta$ parameter on the exchnage-correlation scheme is displayed.
exchange and correlation for the $\beta$ parameter is displayed.
As expected, when increasing the percentage of non-local exchange %
from PBE (0\%), to PBE0 (25\%) up to HF (100\%),
the band gap opens, and the ``degree of localization'' increases
as indicated by the increasing values of $\beta_{\text{max}}$.
This trend is general and also found for the other conjugated polymers 
studied in this work.
A detailed description of the computed band structure for PA in the PA$_{\text{HF}}$, PA$_1$, and PA$_2$ geometries
is reported in Tab.~\ref{tab:betas_pa} for all the methods.

In order to gain more physical insight from our calculations, we have also fitted the data
by using the generalized nearest-neighbors (NN) model presented in Sec.~\ref{sec:method_CBS} and App.~\ref{sec:model}.
This model has three parameters, which approximately correspond to the band gap $E_g$, and the
band widths of the HOMO and LUMO bands (related to the parameters $t_1$ and $t_2$).
The proper relation between the band widths and $t_{1,2}$ is given in 
Eqs.~(\ref{eq:model_bw1}--\ref{eq:model_bw2}).
The main difference between this model and the one introduced in Ref.~[\onlinecite{tomf-sank02prb}] is that
the one used here allows for different band widths for the HOMO and LUMO. While this difference is found
to be small (but generally not negligible) in the cases studied, the generalized model allows for a more accurate fitting
of the electronic structure of polymers. 
As can be seen in Fig.~\ref{fig:PA1_bands}(a-c), the model fitting (dashed, red lines) 
is very accurate for all the
local, semilocal and hybrid functionals. In general the HF data shows the largest
deviation from the model. We believe the non-local nature of the exchange
potential to be the origin of this behavior.

Figure ~\ref{fig:PA1_bands}(d) reports the GW results for PA$_1$. 
The fundamental gap is 0.8 eV at the PBE KS-DFT level, while it increases to
2.05 eV when $G_0W_0$ is applied. This is in very good agreement with previous GW data 
for PA.~\cite{rohl-loui99prl,rohl+01sm,tiag+04prb,varsano_phd}
As already mentioned, GW calculations are performed by using plane-waves and pseudopotentials, while
hybrid-DFT is evaluated on a localized basis set.
In order to assess a possible systematic error due to this procedure, we have also computed the electronic
structure of PA$_{\text{HF}}$ at the LDA, PBE and PBE0 levels, by using the plane-wave implementation of
{\sc Quantum-ESPRESSO}(QE). Results are reported in Tab.~\ref{tab:betas_pa} at the lines X-QE (X=LDA,PBE,PBE0).
The two sets of data are found to be in excellent agreement, allowing for a direct comparison of 
GW and hybrid-DFT data. See also Sec.~\ref{sec:GW_data} and Fig.~\ref{fig:GW_offdiag}(b) for further details.

As a last remark,
according to the standard approach for GW 
calculations,~\cite{hybe-loui86prb1,onid+02rmp} only the Kohn-Sham eigenvalues are corrected, without modifying
the wavefunctions. In other words, only the diagonal matrix-elements of the self-energy 
(on the original DFT Bloch eigenvectors) are considered. The effects of this is analyzed in detail
in Sec.~\ref{sec:GW_data},
when it will be demonstrated that this procedure %
has a sizable effect on the calculation of the complex band structure. For this reason,
the GW-CBS directly computed is reported in a lighter color in the right panel of Fig.~\ref{fig:PA1_bands}(d),
while $\beta_{\text{max}}$ is put in parentheses in Tab.~\ref{tab:betas_pa}.

\subsection{Electronic structure: PPV and PPI}
\label{sec:electronic_structure_applications}

\begin{figure}
   \includegraphics[clip, width=0.45\textwidth]{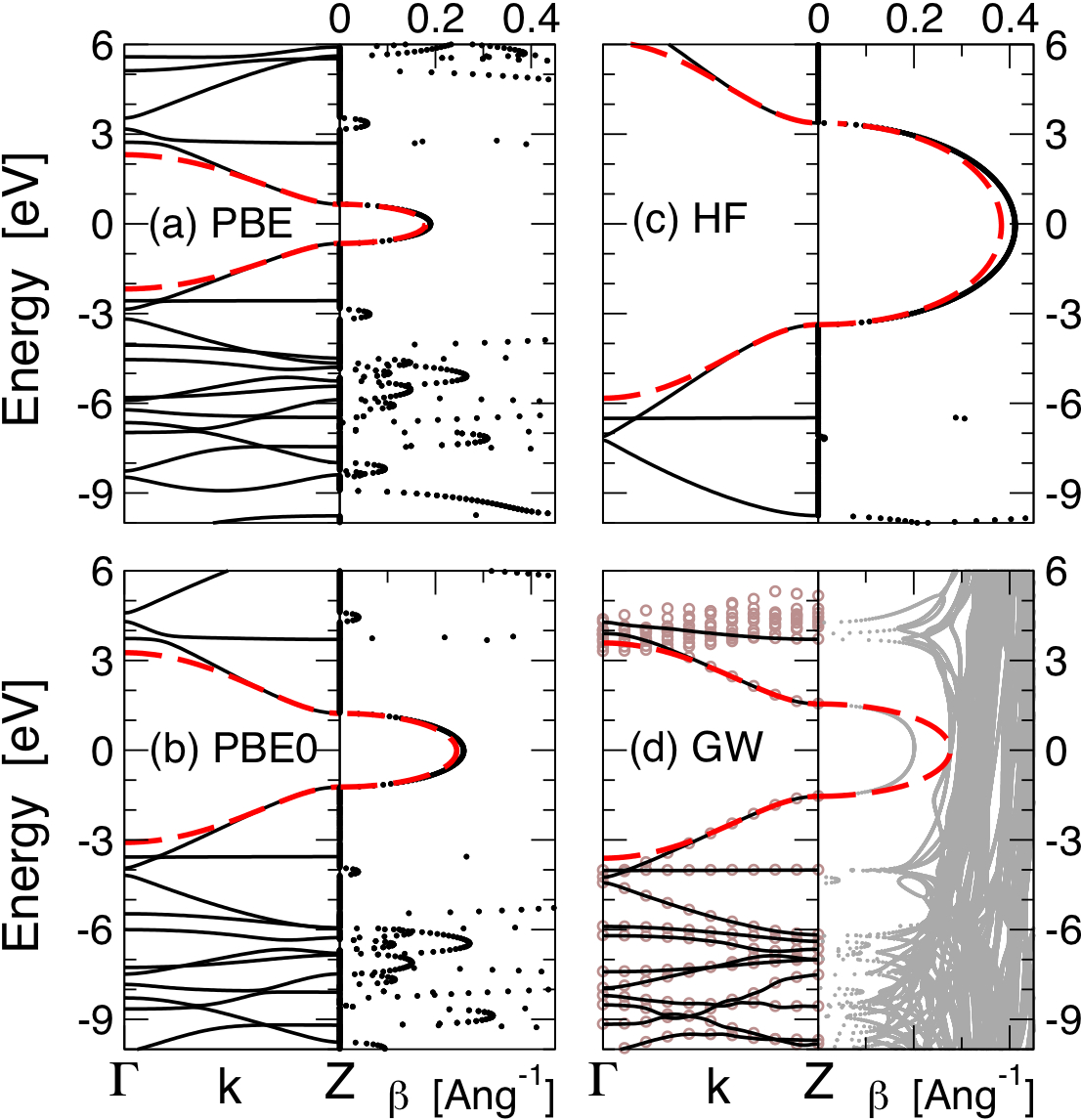}
   \caption{\label{fig:PPV_bands} (color online).
   PPV: Real and complex band structure, 
   as in Fig.~\ref{fig:PA1_bands}.
   }
\end{figure}

\begin{table}
   \centering
   \caption{\label{tab:betas_ppv}
            PPV: Band gap ($E_g$, [eV]), hopping parameters ($t_1$, $t_2$, [eV])
            maximum of $\beta(E)$ [\AA$^{-1}$] inside the gap (computed and fitted from the tight-binding model), 
            as in Tab.~\ref{tab:betas_pa}.
            Last column reports the value $\beta_{\text{max}}$ for the relaxed geometries.
   }
   \begin{ruledtabular} 
   \begin{tabular}{ l | ccc | cc | c }
   \vspace{2pt}
   Scheme &  $\mathbf{E_g}$ & $\mathbf{t_1}$ & $\mathbf{t_2}$ &
             $\boldsymbol{\beta_{\text{max}}}$ & 
             $\boldsymbol{\beta_{\text{max}}}^{\text{model}}$ & 
             $\boldsymbol{\beta_{\text{max}}^{\text{relax}}}$ \\[2pt]
   \hline    
   \hline    
       \multicolumn{7}{c}{} \\
       \multicolumn{7}{c}{ poly {\it para}-phenylene-vinylene (PPV) } \\[2pt]
   \hline
   \vspace{2pt}
   LDA      &   1.28   &  1.07   &    0.016  &    0.19   &  0.18  &  0.18 \\[5pt]
   PW       &   1.31   &  1.07   &    0.017  &    0.19   &  0.18  &  0.19 \\
   BLYP     &   1.31   &  1.07   &    0.016  &    0.19   &  0.18  &  0.19 \\
   PBE      &   1.31   &  1.07   &    0.017  &    0.19   &  0.18  &  0.19 \\[5pt]
   B3PW     &   2.22   &  1.39   &    0.021  &    0.25   &  0.23  &  0.26 \\
   B3LYP    &   2.22   &  1.38   &    0.020  &    0.25   &  0.23  &  0.26 \\
   PBE0     &   2.46   &  1.46   &    0.022  &    0.26   &  0.24  &  0.28 \\[5pt]
   HF       &   6.74   &  2.46   &    0.033  &    0.41   &  0.38  &  0.46 \\[5pt]
   G$_0$W$_0$ &  3.09  &  1.63   &   -0.004  &  0.20(d)  &  0.28  & \\
   \end{tabular}
   \end{ruledtabular}
\end{table}

\begin{figure}
   \includegraphics[clip, width=0.45\textwidth]{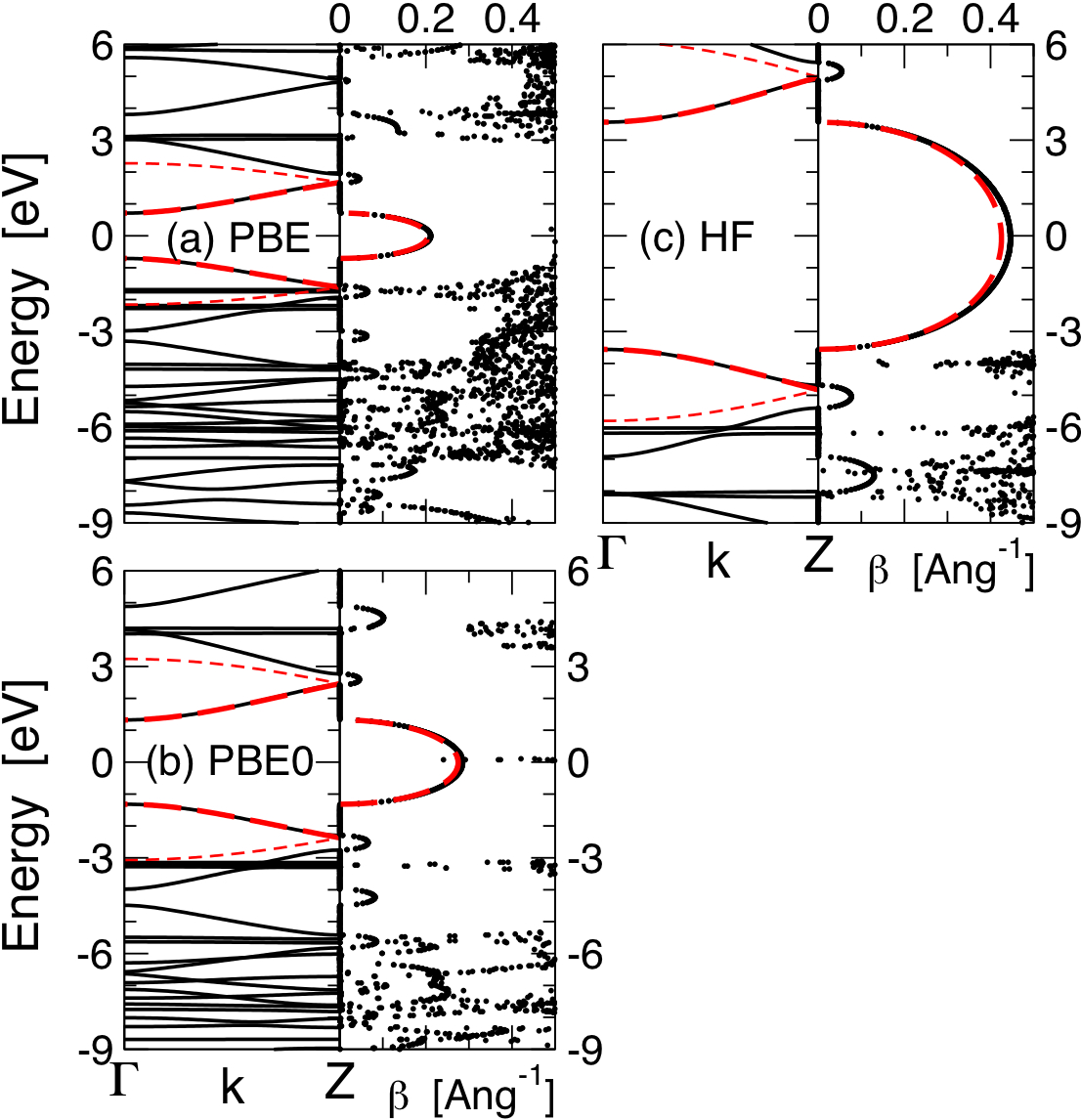}
   \caption{\label{fig:PPI_bands} (color online).
      PPI: Real and complex band structure,
      as in Fig.~\ref{fig:PA1_bands}. The fitting procedure has been applied on a cell with half
      length and then folded, resulting in four bands instead of two.
   }
\end{figure}

\begin{table}
   \centering
   \caption{\label{tab:betas_ppi}
            PPI: Band gap ($E_g$, [eV]), hopping parameters ($t_1$, $t_2$, [eV])
            maximum of $\beta(E)$ [\AA$^{-1}$] inside the gap (computed and fitted from the tight-binding model), 
            as in Tab.~\ref{tab:betas_pa}. Last column reports the value $\beta_{\text{max}}$ for the
            relaxed geometries.
   }
   \begin{ruledtabular} 
   \begin{tabular}{ l | ccc | cc | c}
   \vspace{2pt}
   Scheme &  $\mathbf{E_g}$ & $\mathbf{t_1}$ & $\mathbf{t_2}$ &
             $\boldsymbol{\beta_{\text{max}}}$ & 
             $\boldsymbol{\beta_{\text{max}}}^{\text{model}}$ & 
             $\boldsymbol{\beta_{\text{max}}^{\text{relax}}}$ \\[2pt]
   \hline    
   \hline    
       \multicolumn{7}{c}{} \\
       \multicolumn{7}{c}{ poly-phenylene-imine (PPI) } \\[2pt]
   \hline
   \vspace{2pt}
   LDA      &   1.40   &  1.05   &    0.014  &  0.21 &  0.20  & 0.20  \\[5pt]
   PW       &   1.42   &  1.05   &    0.014  &  0.21 &  0.20  & 0.21   \\
   BLYP     &   1.42   &  1.04   &    0.014  &  0.21 &  0.21  & 0.21   \\
   PBE      &   1.42   &  1.05   &    0.014  &  0.21 &  0.20  & 0.21   \\[5pt]
   B3PW     &   2.38   &  1.36   &    0.020  &  0.27 &  0.26  & 0.29   \\
   B3LYP    &   2.38   &  1.35   &    0.019  &  0.27 &  0.26  & 0.29   \\
   PBE0     &   2.64   &  1.43   &    0.021  &  0.29 &  0.28  & 0.30   \\[5pt]
   HF       &   7.12   &  2.38   &    0.036  &  0.45 &  0.43  & 0.50   \\[5pt]
   \end{tabular}
   \end{ruledtabular}
\end{table}

In this Section we consider two further polymers, namely poly-para phenylene-vinylene (PPV) 
and poly phenylene-imine (PPI).
PPV has been largely investigated for its role in organic (opto-)electronics,~\cite{burr+90nat} while 
oligo phenylene-imine molecules attached to gold leads have been recently considered and the $\beta$
decay coefficents measured.~\cite{choi+08sci,choi+10jacs}
This makes these two polymers particularly appealing for our analysis.

Results for PPV are reported in Fig.~\ref{fig:PPV_bands} and Tab.~\ref{tab:betas_ppv}.
The behavior of the real and complex band structures are qualitatively in agreement with what 
we have found for PA. PBE0 results ($\beta_{\text{max}}$ = 0.26 \AA$^{-1}$) 
are those among the hybrid functionals that best compare with the
interpolated GW data ($\beta_{\text{max}}$ = 0.28 \AA$^{-1}$), though slightly underestimating $\beta_{\text{max}}$ 
and to a larger extent the band gap.
In the case of PPI, results are reported in Fig.~\ref{fig:PPI_bands} and Tab.~\ref{tab:betas_ppi}.
For this polymer, the model fitting has been applied on a cell with half length and then bands have been folded leading
to four interpolated bands instead of two. Despite the reduction of translational symmetry, 
this procedure leads to a better fit because it describes a larger part of the 
frontier electronic structure.
In agreement with previous cases, while we do not have GW results for PPI,
we consider PBE0 data
($\beta_{\text{max}}^{\text{relax}}$=0.29 \AA$^{-1}$) as our best estimate.
In the next Section we will also discuss the comparison of these
computed data with recent experimental results.~\cite{choi+08sci,choi+10jacs}
For the PPV and PPI cases, we have also studied separately the effect of the geometrical
relaxation induced by the different functionals on the CBS.
Such effect is consistent with the trends already observed at fixed geometry.
In the last column of Tabs.~\ref{tab:betas_ppv},~\ref{tab:betas_ppi} we report the $\beta_{\text{max}}$ value
for the relaxed polymer geometries (labelled as $\beta_{\text{max}}^{\text{relax}}$).
The $\beta_{\text{max}}$ parameters increase further with increasing 
fraction of non-local exchange, %
when the geometries are relaxed according to the adopted functional.
While the coupling of the electronic structure with the structural properties is
particularly evident and critical in the case of PA (since the opening of the gap is
due to Peierls distortion of the C-C bonds), it is much less pronounced for PPV and PPI
where it accounts for a correction term only, the leading contribution
being the description of the electronic levels.

\section{Discussion}
\label{sec:discussion}

\subsection{Analysis of the GW data}
\label{sec:GW_data}

In this Section we discuss the effect of some of the approximations involved in the
evaluation of the GW self-energy. In particular we address issues related to the
representation of $\Sigma$ when computing the CBS, as well as the effect of the self-consistency. 
The GW self-consistency is investigated within the static COHSEX approximation.
In doing so, we discuss the localization properties of the resulting Hamiltonians together
with the quality of the NN model fitting of the CBS.
We focus on the case of PA, which is a good prototype for this class of one-dimensional systems.

Let us begin with the representation problem.
As already recalled in Sec.~\ref{sec:GW_interpolation}, 
assuming a large enough subset of Bloch vectors (in principles, all of them), 
the self-energy operator can be represented as:
\begin{equation}
    \Sigma^{\text{GW}}(E) = \sum_{\mathbf{k}} \sum_{mn} \,
                  | \psi_{m\mathbf{k}} \rangle \, \Sigma_{mn}(\mathbf{k}\, E)\, \langle \psi_{n\mathbf{k}} | 
    \label{eq:GW_offdiag}
\end{equation}
[see also Eqs.~(\ref{eq:WF_repres1},\ref{eq:WF_repres2})].
In the usual GW practice, besides evaluating the self-energy by using the underlying KS-DFT
electronic structure for $G$ and $W$ (G$_0$W$_0$ approximation, i.e. no self-consistency),
it is also customary to neglect the off-diagonal band indexes $m\neq n$ in Eq.~(\ref{eq:GW_offdiag}) when 
computing QP energies.
This approximation forces the self-energy to be diagonal on the KS-DFT Bloch states, and thus allows us to modify the
quasi-particle energies without changing the DFT wavefunctions.
If we assume that the correctly represented self-energy [Eq.~(\ref{eq:GW_offdiag})] is
physically short-ranged in real space (further comments follow), as it happens {\it eg} for HF and COHSEX,
when the representation is taken to be diagonal on the Bloch basis 
spurious long range components of the self-energy may
(and typically do) arise, as evident from simple Fourier transform arguments.
The diagonal approximation has been found~\cite{hybe-loui86prb1} to have little effect
on the quasi-particle energy spectrum (at least when LDA wavefunctions are a reasonable starting point).
Our investigations confirm this picture at both the HF and COHSEX level.
As discussed in the following, the case of the complex band structure is more critical.

\begin{figure}
   \includegraphics[clip, width=0.48\textwidth]{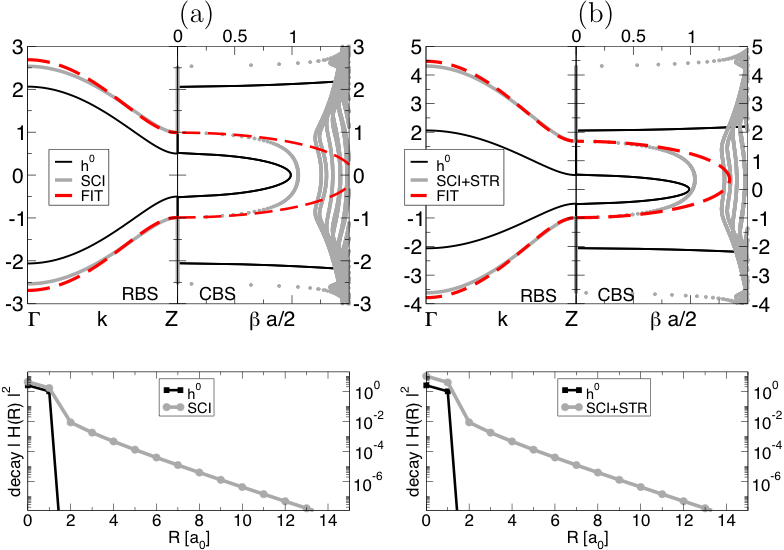}
   \caption{\label{fig:models_scissor} (color online).
   Real and complex band structure for a nearest-neighbors (NN) tight binding model
   Hamiltonian $h^0$ (thin solid black line). A scissor and scissor+stretching corrections to $h^0$
   are applied and shown in panels (a) and (b), respectively. The thick solid gray lines represent the 
   electronic structures obtained while including such corrections.
   The dashed red lines are obtained by a NN tight binding fitting of the 
   real electronic structure after the corrections.
   Lower panels report the spatial decay of the original and corrected Hamiltonians.
   }
\end{figure}

First we focus on a tight-binding model.
In Fig.~\ref{fig:models_scissor} we report the real and complex band structures for such a model, according to
App.~\ref{sec:model}. In order to simulate the effect of a diagonal self-energy, 
we refer to a picture where the $\Sigma$ correction can be modelled as a stretching of the bands (which may be different 
for valence and conduction states) plus a scissor operator applied to the HOMO-LUMO gap.
A simple scissor and a scissor+stretching corrections are applied to the model in Fig.~\ref{fig:models_scissor}(a,b)
(solid black lines for the original model, thick light gray lines for the corrected ones).
The new Hamiltonian including the corrections is now longer ranged than the original nearest neighbors Hamiltonian
$h^0$, because of the
non-local projectors used to express the scissor and scissor+stretching corrections.  
The different spatial decay of the pristine and corrected Hamiltonians 
is shown in the lower panels of Fig.~\ref{fig:models_scissor}.
We then fit the corrected Hamiltonian by using again the NN model (dashed, red lines). This fitting 
emulates 
the real band structure, but using a short ranged (nearest-neighbors) Hamiltonian.
The effect on the CBS is evident and sizeable. The simply shifted and stretched 
electronic structures leads to little corrections
to the CBS, while much bigger corrections are obtained when considering the fitted short-ranged Hamiltonians.  
Because CBS measures the decay of the evanescent states in real space, it is not surprising that
a method which does not update the wavefunctions (as the diagonal self-energy corrections) is not able to
capture all of the physics involved in the change of the electronic structure.
\begin{figure}
   \includegraphics[clip, width=0.45\textwidth]{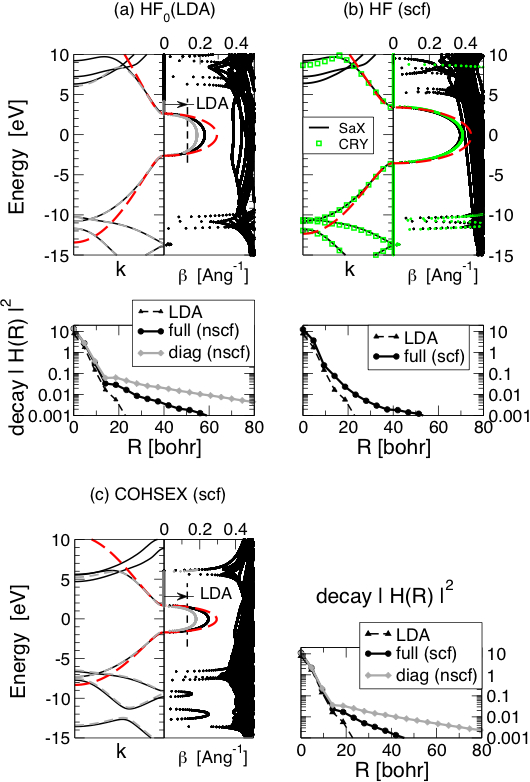}
   \caption{\label{fig:GW_offdiag} (color online).
   Real and complex band structure for PA$_\text{HF}$ at the HF and COHSEX levels. 
   Panel (a) shows non-self-consistent HF results 
   with a diagonal (gray lines) and fully off-diagonal (black lines) representation of the correction.
   Panel (b) reports the same data for the self-consistent HF solution, as computed from SaX
   (solid black lines) and CRYSTAL (green triangles).
   Panels (c) shows COHSEX data: Diagonal non-self-consistent (gray lines) and fully self-consistent 
   (black lines) data are reported.
   In both cases, a NN tight-binding fit of the real and complex band structures
   is performed (dashed red lines).
   Lower panels show a measure of the spatial decay of the COHSEX and HF Hamiltonian matrices on the WF basis.
   }
\end{figure}

In order to further numerically support this interpretation and to investigate the effect of the
self-consistency, we have evaluated the HF and COHSEX self-energies for PA$_\text{HF}$.
At first we have done so non-self-consistently (self-energies evaluated on the LDA wavefunctions),
with and without the diagonal approximation. Then self-consistent~\cite{note_COHSEX} COHSEX results are provided. 
HF results are plotted in Fig.~\ref{fig:GW_offdiag}(a,b), COHSEX data in panel (c).
LDA $\beta_{\text{max}}$ is shown with a dashed line in the CBS panels as a reference.
Regarding the real band structure, we found that the inclusion of off-diagonal matrix elements is not very relevant for PA, 
the bands being almost overlapping for both HF and COHSEX [diagonal data shown in panels (a,c) by dashed gray lines]. 
The situation is different for the CBS, as it is highlighted in Fig.~\ref{fig:GW_offdiag}(a).
The HF$_0$ correction of $\beta_{\text{max}}$ from the full off-diagonal representation is 
almost twice as big as the diagonal correction. 
This confirms the behavior observed with the models in Fig.~\ref{fig:models_scissor}.

This observation also correlates with the decay of the HF$_0$(LDA) Hamiltonian
reported in the lower part of panel (a). 
As for the models [Fig.~\ref{fig:models_scissor}], the diagonal representation
induces a much longer (and unphysical) decay. The same situation is found for COHSEX (the off-diagonal results at the
first SCF iteration are not shown). The proper Hamiltonian decay (black line, circles) is clearly longer
ranged than the LDA results, because of the non-local contribution of the exchange operator.
The decay of the exchange potential is driven by that of the density matrix, which in turn is 
related~\cite{he-vand01prl} to $\beta$ at the branch point. Being $\beta_{\text{max}}$ typically underestimated at the 
LDA level, so is the decay of the HF$_0$ Hamiltonian. The effect of self-consistency of HF (as well as COHSEX)
is then to reduce such over-delocalization and to produce shorter ranged self-energies. This is shown in the decay plot of
panels (b,c). 

This behaviour has strong consequencies regarding the quality of the NN model fit.
In the case of local and self-consistent hybrid functional calculations from {\sc CRYSTAL}, the model fit
(dashed red lines in Figs.~\ref{fig:PA1_bands},~\ref{fig:PPV_bands}) 
works very well when compared with the full calculations,
despite its semplicity. This is not the case when comparing with the diagonal corrections of 
Fig.~\ref{fig:GW_offdiag}(a,c). The off-diagonal representation improves the situation but does not solve the problem. 
The failure of the model fit in Fig.~\ref{fig:GW_offdiag}(a) is in fact mostly related to the 
decay of the exchange operator. A much better fit is then found when full self-consistency
is included, as in Fig.~\ref{fig:GW_offdiag}(b,c).
Finally, we have also compared the HF results from SaX (plane-waves and pseudopotentials) 
and CRYSTAL (localized basis, full electron). Results are reported in Fig.~\ref{fig:GW_offdiag}(b) 
by the black solid lines and the green triangles, respectively. 
We find excellent agreement for both the real and complex band structures, 
confirming that the results from the two codes are well comparable and free of any systematic error.

In the light of the discussion above,
when presenting the diagonal G$_0$W$_0$ data for PA and PPV 
[Figs.~\ref{fig:PA1_bands}(d),\ref{fig:PPV_bands}(d)], the CBS is better described by
the interpolated data (dashed lines) instead of that directly calculated 
from the diagonal GW corrections (solid thin lines).
Similar conclusions about the importance of describing changes to wavefunctions
when applying GW to transport calculations are reported also in Refs.~[\onlinecite{rang+11prb, stra+11prb, tamb+11prb}].

\subsection{Electronic structure}
Now we can turn to discussing the accuracy of the electronic structure calculations
for conjugated polymers.
First of all we note that our results for PA$_1$ and PPV are in good agreement with previously 
published~\cite{rohl-loui99prl,rohl+01sm,tiag+04prb,varsano_phd} G$_0$W$_0$ results.
In particular, for PA$_1$ we obtain a GW gap $E_g = 2.05$ eV, to be compared with 2.1 eV 
(Ref.~[\onlinecite{rohl-loui99prl},\onlinecite{rohl+01sm}]) and 2.13 eV (Ref.~[\onlinecite{varsano_phd}]).
These results are obtained for isolated chains of PA. In the case of a crystal, the
gap shrinks\cite{tiag+04prb} to 1.8 eV
due to interchain interactions.
For the isolated chain of PPV, Rohlfing {\it et al}.~\cite{rohl-loui99prl,rohl+01sm} found $E_g = 3.3$ eV (G$_0$W$_0$),
which compares reasonably well with our G$_0$W$_0$ result of 3.09 eV [see Tab.~\ref{tab:betas_ppv}].
In general, the overall shape (including the band widths) 
of the GW-corrected band structures for PA and PPV computed in this work 
is in excellent agreement with that of Ref.~[\onlinecite{rohl-loui99prl}].

From a methodological point of view,
the accuracy of G$_0$W$_0$ corrections (based on the LDA electronic structure) for organic molecules
has been recently widely addressed,~\cite{tiag+08jcp,rost+10prb,kaas-thyg10prb,blas+11prb,fabe+11prb}
comparing with different implementations of self-consistent GW and experimental results.
G$_0$W$_0$(LDA) is found to underestimate ionization potentials more than
in extended systems, suggesting that a certain degree of self-consistency tends
to improve on the results. Moreover, self-consistency is found to further lower the
HOMO level and increase the fundamental gap. This leads to larger estimates
of $\beta_{\text{max}}$. 

In terms of a direct quantitative comparison with experiments, some issues have
to be taken into account. First, electronic structure (and optical) measurements for polymers 
typically distinguish between crystalline grains and amorphous regions. Isolated chains
are considered to resemble more (and to be used as rough models for) the amorphous regions. %
Clearly, a direct theory-experiment
comparison may suffer from systematic errors (interchain interactions, medium polarization, electrostatic effects).
These features generally tend to reduce the fundamental gap
wrt that of the ideally isolated chain.
With this in mind, we can compare data calculated here with experimental data from 
photoemission (PES) or scanning tunneling (STS) spectroscopies.
Rinaldi et al., measured~\cite{rina+01prb} the electronic gap of PPV films (on a GaAs substrate) by means
of STS. They were able to estimate $E_g \sim 3$ eV.
Kemerik et al.~\cite{keme+04prb} used also STS and
found the fundamental gap of PPV [film deposited on Au(111)] to be around 2.8 eV.
All of these results are to be reasonably considered as lower bounds of the theoretical gap for the isolated PPV chain.

\subsection{Complex band structure: trends}
As can be directly inferred from the model described in App.~\ref{sec:model},
as well as from Ref.~[\onlinecite{tomf-sank02prb}], 
the important parameters that determine the behavior of $\beta(E)$ (and $\beta_{\text{max}}$) 
are the band gap $E_g$, and the effective band
widths of the states around the gap, given {\it eg} in terms of the hopping $t_1$.
While our model (see App.~\ref{sec:model}) includes a second parameter $t_2$ to 
describe the difference in the band widths of the frontier bands, 
considering that the ratio $t_2/t_1$
ranges from 0.1 to 0.01 or less, corrections to the model due to
$t_2$ are not particularly relevant for the cases studied here.
According to Eqs.~(\ref{eq:betamax_model},\ref{eq:betamax_model2}), $\beta_{\text{max}}$ is mostly 
determined by the $E_g/t_1$ ratio.
Even though this is just a simplified NN model, our numerical investigations suggest that the model is widely
applicable (using a folding technique as in the case of PPI when needed).

In general the band gaps are expected to increase with the fraction of non-local exchange included
in the hybrid functional. For covalently bonded systems, the same trend is expected for the band widths.
Numerically, in all our examples, while increasing the energy gap, HF increases also the band widths. 
The same trend is found for all the hybrid functionals we have investigated. Since both $E_g$ and $t_1$ increase, 
it is not trivial to understand {\it a priori} which mechanisms would dominate. Indeed, it is very clear
that the band gap opens more than the band width, leading to a clear trend of $\beta_{\text{max}}$ increasing
when a larger fraction of exchange is included in the calculation.
The same trend is also found for the GW results, even though we had to extrapolate the CBS from the real band structure 
by using the model fitting (as described in details in Sec.~\ref{sec:GW_data}).

\begin{figure}
   \includegraphics[clip,width=0.45\textwidth]{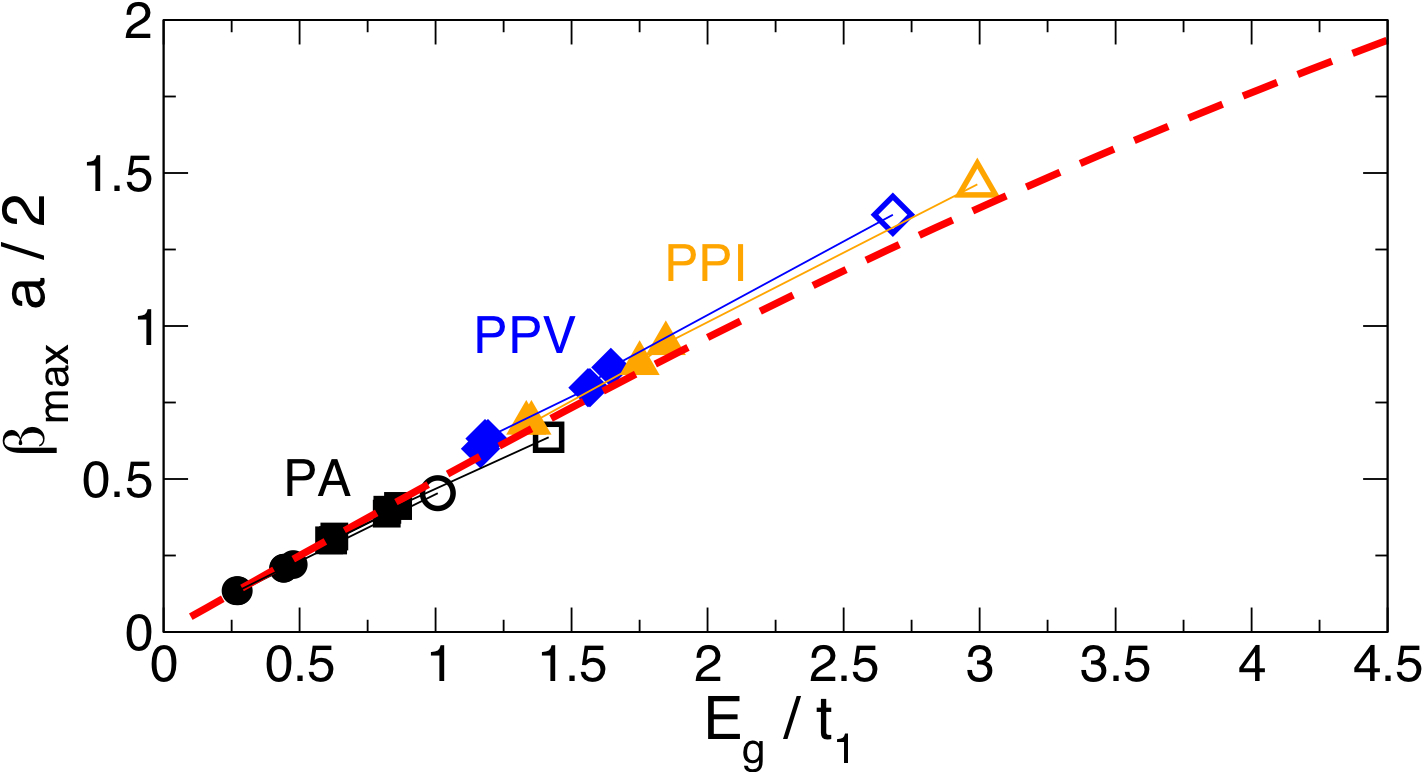}
   \caption{ \label{fig:model_fitting} (Color online). Computed $\beta_{\text{max}} a/2$ versus
             $E_g / t_1$, $E_g$ being the band gap, $t_1$ the
             effective hopping, and $a$ the lattice parameter.
             All the polymers and XC functionals or methods are plotted.
             Black circles: PA$_1$, squares: PA$_2$; blue diamonds: PPV; orange triangles: PPI.
             Open symbols refer to HF results.
             Dashed red line: $\beta_{\text{max}}$ according to the 
             tight-binding model.
            }
\end{figure}
In Fig.~\ref{fig:model_fitting} we report a synthetic view of all the computed values of 
$\beta_{\text{max}}$ (times $a/2$, $a$ being the polymer lattice parameter), including all
electronic structure methods for PA (PA$_1$, PA$_2$ are shown; PA$_{\text{HF}}$ is not shown because almost overimposed
to the other PA geometries), PPV and PPI.
Those data are plotted against the $E_g/t_1$ value. The ideal curve from the
tight-binding NN model is reported (dashed red line).
The agreement between the computed and the modelled data is remarkable for all the cases studied.
HF data are reported with empty symbols, showing in general a slightly worse agreement 
(as already discussed in Sec.~\ref{sec:electronic_structure_applications}).
On the basis of the above relations, and according to the results reported in 
Fig.~\ref{fig:model_fitting}, 
we suggest the use of the model fitting to extrapolate information 
about $\beta_{\text{max}}$ from experiments able to investigate the electronic structure.
This would allow for an indirect measure of the CBS and the related parameters (as $\beta_{\text{max}}$).
\subsection{Comparison with transport data}

As discussed in the introduction, in order to compute
the $\beta$ decay of the current (or conductance) in a metal-insulator-metal (MIM) junction, 
we need two different ingredients: ($i$) the knowledge of the CBS
for the infinitely long insulating region, and ($ii$) the position of the Fermi level of the MIM junction
wrt the band structure of the insulator. This is depicted in Fig.~\ref{fig:cartoon}.
Assuming that at low bias the current is carried by the states close to the Fermi level of the junction, 
we basically need to know how these states decay into the insulating region (the polymer, in the present case).
Once the CBS is known, either the Fermi level alignment is computed explicitly or it is estimated on the basis of 
physical considerations. While the direct calculation is feasible (but demanding) and in some cases necessary, 
it is also possible to give a first estimate of the Fermi level according to the so-called MIGS
(metal-induced gap states) theory.~\cite{ters84prl,ters84prb,tomf-sank02prb}
As proposed by Tersoff,~\cite{ters84prl} if the metal DOS around the Fermi level is sufficiently featureless and the
MIGS penetrate deep enough in the insulating region, 
the Fermi level of the metal-insulator junction is approximately pinned
at the charge-neutrality level (often close to the midgap point) in order to avoid charge imbalance at the interface.
The charge neutrality level can be easily identified from the CBS, as the energy where $\beta(E)$ reaches
its maximum inside the gap. From this perspective, $\beta_{\text{max}}$ is a first estimate of the
experimental $\beta$ decay. In general, the Fermi level will move from the charge neutrality level, and the $\beta$
decay will change accordingly.
The physical reasons leading the Fermi level to shift are mostly related to
the charge transfer (and dipole formation) at the interface. This explains why different chemical linking groups on the
same molecule may lead to different values of $\beta$. 
In general, the $\beta_{\text{max}}$ value computed from the CBS may be regarded as an upper bound for $\beta$
and thus as a lower bound for the ability of the insulating layer to allow the current to tunnel through
the junction.

The above discussion stresses the fact
that the experimentally measured $\beta$ does not in general depend~\cite{tomf-sank02prb} 
only on the electronic structure of the insulator (especially the CBS), but also on the details of the interface 
(which determines the position of the Fermi level). 
For instance, recent measurements on alkanes (oligomers of PE) determined a $\beta$ decay length of 0.71--0.76 \AA$^{-1}$ for
for -NH$_2$ terminations,~\cite{venk+06nl,park+09jacs,hybe+08jpcm} while it has been found 
in the range 0.8--0.9 \AA$^{-1}$ (and more disperse) for thiolated molecules.~\cite{holm+01jacs,xu-tao03sci,wold+02jpcb,hybe+08jpcm}
Recent calculations confirm this picture also for conjugated
polymers connected to gold leads through different chemical groups.~\cite{peng+09jpcc}
In that work, the Authors have studied a number of oligomers including oligo-phenylenes 
(whose infinite polymer is poly-para-phenylene, PPP). 
In the case of the molecule connected to gold leads via thiol groups,
after investigating the interface-DOS projected on the molecule, they find that the Fermi level aligns close to
the midgap point (reasonably the charge neutrality level considered here), off by few tenths of eV ($\sim$0.2-0.3 eV, 
the molecular gap being about 2 eV) at the LDA level.
Since the basic unit of PPP (a phenyl ring) is the same as part of the monomers forming PPV and PPI, we assume the
charge neutrality condition to be almost fulfilled if we were considering Au-PPV-Au and Au-PPI-Au junctions with thiol
anchoring. We can thus estimate $\beta_{\text{max}}$ to be a good estimate of $\beta$, keeping in mind that the a small
deviation from midgap would slightly decrease $\beta$.

Recent experiments~\cite{choi+08sci,choi+10jacs} reported $\beta$ = 0.3 \AA$^{-1}${} for oligomers of PPI connected to 
gold through thiols. This number is in very good agreement with the value $\beta_{\text{max}}$ = 0.29 \AA$^{-1}${} we found
for PPI using PBE0 (see Tab.~\ref{tab:betas_ppi}). According to our findings for PA and PPV 
(see Tabs.~\ref{tab:betas_pa},\ref{tab:betas_ppv}), GW should give results for $\beta$ comparable to PBE0, 
but slightly larger. 
In order to compare with other existing (experimental and theoretical) 
results for oligo-phenylenes (PPP in the infinite limit)~\cite{wold+02jpcb,kaun+03prb,venk+06nat,quek+09nl} 
we would need to address separately the issue of the Fermi level alignment (tending to decrease $\beta$ wrt the ideal 
$\beta_{\text{max}}$) and that of the phenyl twist-angle (which goes in the direction of increasing $\beta$).
This will be the subject of future work.
Coming to the case of PA,
we note that recent experimental results~\cite{he+05jacs,meis+11nl} report $\beta$-values
of 0.22 \AA$^{-1}${} for molecules similar to oligo-acetylenes. While this number is in very good agreement
with our GW (and PBE0) results for PA$_1$ (and consistent with PA$_\text{HF}$, see Tab.~\ref{tab:betas_pa}), 
the large variability of $\beta$ with the structural parameters of PA does not allow us
to be conclusive on the assessment of the theory vs the experiment for this specific case. 
Nevertheless, our findings suggest that the use of PBE0 or GW results, together with a proper determination of the
Fermi level alignment, will provide a reasonable approximation. 
We also note that gap underestimation and {\it a priori} assumption of the validity of the MIGS theory
tend to partly cancel each other, and fortuitous agreement of experimetally
measured $\beta$ values with LDA calculations may occur.

\section{Conclusions}
\label{sec:conclusions}

In this work we have computed from first principles the real and complex band-structure
of prototype alkylic and conjugated polymer-chains using a number of theoretical schemes, 
ranging from local and semilocal to hybrid-DFT,
and GW corrections. The accuracy of these different methods has been evaluated and compared with
existing theoretical and experimental data, both in terms of the electronic structure and transport properties.
From the CBS 
the $\beta$ decay parameter, which governs 
non-resonant tunneling experiments through metal-insulator-metal junctions, can be computed.

In doing so we have stressed the formal analogy of hybrid-DFT and GW (especially in the COHSEX formulation), 
and the interpretation of the hybrid-DFT electronic structure as an approximation to the proper quasi-particle
spectrum. We have also described in detail how to interpolate GW results by using a Wannier function scheme.
In this case we have found that while the real band structure is always well interpolated, the CBS needs the
self-energy real-space decay to be properly treated (off-diagonal representation and self-consistency of the
wavefunctions). 

We have numericaly investigated four polymers, namely poly-ethylene (PE), poly-acetylene (PA), 
poly-para-phenylene-vinylene (PPV), and poly-phenylene-imine (PPI). Our results compare well with the existing
theoretical and experimental literature. Among the hybrid functionals studied, PBE0 results 
compare best with the G$_0$W$_0$ electronic structure. While the band gaps may still have 
a non-negligible deviation from GW, the agreement is remarkable on the CBS and $\beta$ coefficient.
The comparison with transport data (when available) is also very promising.
This suggests PBE0 as an efficient and reliable alternative to GW for these class systems, at least for
transport properties. More generally, a
systematic application of hybrid functionals to improve the accuracy of DFT-based electronic structure results is appealing,
while further developments along the lines of Ref.~[\onlinecite{marq+11prb}] are probably needed.

\section{Acknowledgements}

We thank D. Varsano, P. Bokes, N. Marzari, I. Dabo, X.-F. Qian, and S. de Gironcoli for
fruitful discussions.
This work made use of the high performance computing facilities of CINECA,
Rutherford Appleton Laboratory (RAL), Imperial College London,
and -- via membership of the UK's HPC Materials Chemistry Consortium
funded by EPSRC (EP/F067496) --
of HECToR, the UK's national high-performance computing service,
which is provided by UoE HPCx Ltd at the University of Edinburgh, Cray Inc and NAG Ltd,
and funded by the Office of Science and Technology through
EPSRC's High End Computing Programme.
This work was supported in part by Italian MIUR, through
Grant No. PRIN-2006022847 and FIRB-RBFR08FOAL\_001.

\appendix

\def\Xint#1{\mathchoice
{\XXint\displaystyle\textstyle{#1}}%
{\XXint\textstyle\scriptstyle{#1}}%
{\XXint\scriptstyle\scriptscriptstyle{#1}}%
{\XXint\scriptscriptstyle\scriptscriptstyle{#1}}%
\!\int}
\def\XXint#1#2#3{{\setbox0=\hbox{$#1{#2#3}{\int}$ }
\vcenter{\hbox{$#2#3$ }}\kern-.5\wd0}}
\def\ddashint{\Xint=}
\def\dashint{\Xint-}

\section{One-dimensional model}
\label{sec:model}

\begin{figure}
   \includegraphics[clip, width=0.40\textwidth]{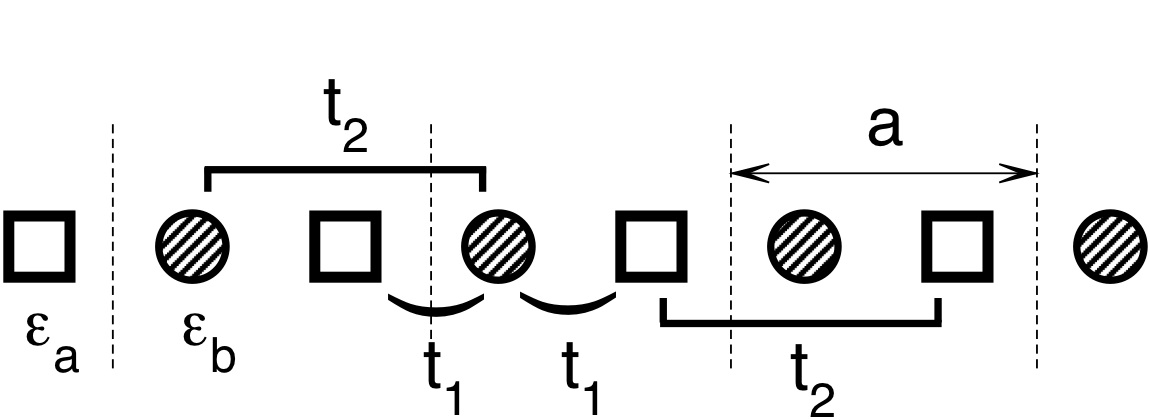}
   \caption{\label{fig:model} Cartoon of the Hamiltonian adopted
   to fit the computed data, according to Eq.~(\ref{eq:model_3p}).
   }
\end{figure}

In the case of one dimensional systems like conjugated polymers, 
numerical results for $\beta$
can be rationalized in terms of a simple tight binding model 
as presented in Ref.~[\onlinecite{tomf-sank02prb}].
In their work, Tomfohr and Sankey presented a two-band
model which provides an analytical expression for the complex band structure (CBS)
within the fundamental energy gap. This model can also be used to fit the real band structure
of realistic systems in order to evaluate the CBS analytically.
The model is described in terms of two inequivalent sites ($\epsilon_{a,b}$) with
nearest-neighbors (NN) hopping $t_1$. These three parameters can be recast into $E_g$ (the fundamental gap), $t_1$ and
a further shift of the energy levels, which has no physical meaning. Moreover, it is shown that
in this case $\beta_{\text{max}}$ (the maximum value of $\beta(E)$ within the fundamental gap) 
depends only on $E_g / t_1$, according to: 
\begin{eqnarray}
\label{eq:betagamma}
   \beta(E) \, a/2       &=& \text{ln}\left[ \gamma(E) + \sqrt{ \gamma(E)^2 -1} \right], \\
   \gamma(E)             &=& \frac{(E-E_v)(E_c-E)}{2t_1^2} +1, \\
    E_{\beta_{\text{max}}} &=& \Sigma = \frac{E_v + E_c}{2} \\
   \gamma(E_{\beta_{\text{max}}}) &=& 1 + \frac{1}{8} \left( \frac{E_g}{t_1} \right)^2.
\label{eq:betamax_model}
\end{eqnarray}
All the details are given in Ref.~[\onlinecite{tomf-sank02prb}].
We note however that this model is unable to reproduce any difference in the widths of the two
bands (thus resulting in the CBS maximum being located at mid-gap), 
while in our realistic simulations we typically find the LUMO
bandwidth to be somewhat greater than that of the HOMO.
We have then generalized the above to the following three-parameters model, which is more suitable to determine
the physical relevant quantities from our simulations.
Extending the previous model, we add second NN interactions (with strength $t_2$) between 
equivalent sites, $t_1$ being the hopping between inequivalent sites as in the previous model.
The model Hamiltonian is the following:
\begin{eqnarray}
\nonumber
  H &=& \sum_{\mathbf{R}} \, \epsilon_{a} 
       \psi^{\dagger}_{a,\mathbf{R}} \psi^{\phantom{\dagger}}_{a,\mathbf{R}} + 
       \sum_{\mathbf{R}} \, \epsilon_{b} 
       \psi^{\dagger}_{b,\mathbf{R}} \psi^{\phantom{\dagger}}_{b,\mathbf{R}} + \\
\nonumber
    & & \sum_{\mathbf{R}} \, t_1 \left[ 
          \psi^{\dagger}_{a,\mathbf{R}} \psi^{\phantom{\dagger}}_{b,\mathbf{R}} 
       +  \psi^{\dagger}_{a,\mathbf{R}+1} \psi^{\phantom{\dagger}}_{b,\mathbf{R}} 
       \right] + \text{cc} + \\
    & & \sum_{\mathbf{R}} \, t_2 \left[ 
         \psi^{\dagger}_{a,\mathbf{R}+1} \psi^{\phantom{\dagger}}_{a,\mathbf{R}} 
       + \psi^{\dagger}_{b,\mathbf{R}+1} \psi^{\phantom{\dagger}}_{b,\mathbf{R}} \right] 
       + \text{cc},
\label{eq:model_3p}
\end{eqnarray}
where $a,b$ indicate inequivalent sites and $R$ is a cell index. 
The model is pictorially described in Fig.~\ref{fig:model}.
The parameters $t_1$, $t_2$ are in general complex numbers.
Taking $t_1$, $t_2$ to be real for semplicity, 
the analytical expressions of the energy bands read:
\begin{equation}
  E_{1,2}(k) = \Sigma + 2t_2 \, x(k) \pm \left[ \Delta^2 + 2t_1^2 \, x(k) \right]^{\frac{1}{2}},
\end{equation}
where, assuming $\epsilon_a > \epsilon_b $, %
we have set:
\begin{eqnarray}
   E_v &=& \epsilon_b -2t_2 ; \qquad E_c = \epsilon_a -2t_2 ; \\
   x(k) &=& 1 + \text{cos}(ka) ; \\
   \Sigma &=& \frac{1}{2} \left( E_c +E_v \right) ; \\
   \Delta &=& \frac{1}{2} \left( E_c -E_v \right) = \frac{E_g}{2} . 
\end{eqnarray}
In this picture $E_{c,v}$ are the onsets of the valence and conduction bands, $a$ is
the lattice parameter, $k$ runs from $-\frac{\pi}{a}$ to $\frac{\pi}{a}$.
Under the condition $|t_2| \ll |t_1|$, the band gap of the model is still at the Brillouin zone edge.
A differenc choice of the relative phases of the model parameters would be needed
to have the band gap at $\Gamma$ (which is the case for PPI).

It is also possible to express the results of the model in terms of more physical parameters
such as the energy gap and the HOMO and LUMO band widths;
\begin{eqnarray}
   \label{eq:model_bw1}
   W_c = \left[ \Delta^2 + 4 t_1^2 \right]^{1/2} -\Delta + 4 t_2 \\
   \label{eq:model_bw2}
   W_v = \left[ \Delta^2 + 4 t_1^2 \right]^{1/2} -\Delta - 4 t_2
\end{eqnarray}
Because in the realistic calculations the bands of interest may cross other bands far 
from the $k$ corresponding to the gap, we consider partial ($\widetilde{W}_c$, $\widetilde{W}_v$) 
and not full band widths. These quantities are defined by
the amplitude of the bands in a limited range of the BZ around the fundamental gap.
This yields a much better agreement of the model bands with the calculated bands close to
$E_g$, which is the energy range of interest.
In order to extract the parameters $E_g, t_1, t_2$ from our calculations we used the following
relations (under the restriction that the band gap is direct 
and located at the Brillouin zone edge $k=\pi/a$):
\begin{eqnarray}
    \label{eq:model_k0}
    x_0   &=& x(k_0), \\
    \label{eq:model_t2}
    t_2   &=& \frac{1}{4x_0} \left(\widetilde{W}_c -\widetilde{W}_v \right), \\
    t_1^2 &=& \frac{1}{2x_0} \left( (\widetilde{W}_c -2t_2\, x_0 + \Delta)^2 -\Delta^2 \right) .
    \label{eq:model_t1}
\end{eqnarray}

Following Ref.~[\onlinecite{tomf-sank02prb}], 
once we have parametrized the model, we can give an analytical expression for the CBS, which
in our case reads:
\begin{eqnarray}
   \gamma(E)  = \frac{(E-E_v)(E_c-E)}{2( t_1^2 + 2 Et_2 -2t_2\Sigma)} +1, 
\end{eqnarray}
where Eq.~(\ref{eq:betagamma}) connecting $\beta$ to $\gamma$ holds unchanged.
The maximum of $\gamma(E)$ can be found analytically, and the resulting expression can be further
simplified under the assumption $|t_2| \ll |t_1| $;
\begin{eqnarray}
    E_{\beta_{\text{max}}}        &\sim& \Sigma - t_2 \left( \frac{\Delta}{t_1} \right)^2, \\
   \gamma(E_{\beta_{\text{max}}}) &\sim& 1 + \frac{1}{8} \left( \frac{E_g}{t_1} \right)^2 \, 
                                      \left[ 1 + \frac{t_2^2}{t_1^2} \, \left( \frac{\Delta}{t_1} \right)^2 \right].
\label{eq:betamax_model2}
\end{eqnarray}
\section{Computational details}
\label{sec:details}

The {\sc CRYSTAL09} software package~\cite{CRYSTAL09} 
perform calculations based on the expansion of
the crystalline orbitals as a linear
combination of a local basis set consisting of atom centered Gaussian orbitals.
A 6-31G* contraction double valence (one $s$, two $sp$ and one $d$ shells) quality basis sets 
have been selected to describe carbon and nitrogen atoms;
the most diffuse $sp$ ($d$) exponents are $\alpha^{\rm C}=0.1687\ (0.8)$ and
$\alpha^{\rm N}=0.2120(0.8)$ Bohr.$^{-2}$
The hydrogen atom basis set consists of a 31G* contraction (two $s$, one $p$ shells):
the most diffuse $s$ and $p$ exponents are 0.1613 and 1.1 Bohr.$^{-2}$
The self consistent field procedure was converged to a tolerance in
the total energy of $\Delta E=2 \cdot 10^{-7}$ Ry per unit cell.~\cite{note-details-crystal}
Reciprocal space sampling was performed on Monkhorst-Pack grid with
12 $\mathbf{k}$-points. 
The thresholds for the maximum and RMS forces
(the maximum and the RMS atomic displacements) 
have been set to 0.00090 and 0.00060 Ry/Bohr
(0.00180 and 0.00120 Bohr).
The calculations performed with {\sc Quantum-ESPRESSO} (QE) adopt a plane waves basis set and norm-conserving
pseudopotentials to describe the ion-electron interaction. The kinetic energy
cutoff has been set to 45 Ry for wavefunctions. For ionic relaxation, total energy LDA and GGA calculations 
use a Monkhorst-Pack grid of  8, 6, and 6 $\mathbf{k}$-points for PE, PPV and PPI respectively (PA geometries are taken
from the literature and not relaxed with QE).
The convergence threshold on the atomic forces
has been set to $10^{-3}$ Ry/Bohr.
A minimum distance of 20 Bohr between chain replica is used.

When performing GW calculations using {\sc SaX}, the $\mathbf{k}$-points
grids have been made finer by using 50, 50 and 20 $\mathbf{k}$-points
for PA, PE and PPV, respectively. 
The long-range divergence of exchange-like Coulomb
integrals is treated using a generalized version~\cite{mart+11tobe} of the
approach given by Massidda et al.~\cite{mass+93prb}
The same approach has
also been used when performing hybrid-DFT calculations with QE.
Note that other schemes to treat exchange in one-dimensional
systems have been proposed.~\cite{isma06prb,rozz+06prb,mari+09cpc,li-dabo11prb}
The Godby-Needs plasmon-pole model~\cite{godb-need89prl} has been used, setting the fitting energies
at 0.0 and 2.0 Ry along the immaginary axis, for all the cases.
A kinetic energy cutoff of 6 Ry has been used to represent the polarizability $P$
and the dynamic part of the screened Coulomb interaction $W$ on a plane-wave basis, while 
a cutoff of 45 Ry has been used for the exchange operator.
In order to converge the sums over empty states for the polarizability (self-energy), a total number of 
288, 288, 288 (288, 288, 608) states has been used for PA, PE, and PPV respectively.
This corresponds to an equivalent transition-energy cutoffs of 53, 53, 35 eV (53, 53, 44 eV).
Interchain distance has been increased to $\simeq$30 Bohr to control spurious interactions
of periodic replica.
The QP corrections are computed by evaluating the diagonal matrix elements of the 
self-energy operator $\langle n\mathbf{k} | \Sigma_{\text{xc}} | n\mathbf{k} \rangle$,
unless explicitly stated.


\begin{thebibliography}{100}

\bibitem{nitz-ratn03sci}
A.~Nitzan and M.~A. Ratner,
\newblock Science {\bf 300}, 1384 (2003).

\bibitem{floo+04sci}
A.~Flood, J.~Stoddart, D.~Steuerman, and J.~Heath,
\newblock Science {\bf 306}, 2055 (2004).

\bibitem{agra+03prep}
N.~Agra{\"\i}t, A.~L. Yeyati, and J.~M. {van Ruitenbeek},
\newblock Phys.~Rep.~ {\bf 377}, 81 (2003).

\bibitem{bour01book}
J.~P. Bourgoin,
\newblock Lecture notes in physics {\bf 579}, 105 (2001).

\bibitem{holm+01jacs}
R.~Holmlin, R.~Haag, M.~Chabinyc, R.~Ismagilov, A.~Cohen, A.~Terfort, M.~Rampi,
  and G.~Whitesides,
\newblock J. Am. Chem. Soc. {\bf 123}, 5075 (2001).

\bibitem{chab+02jacs}
M.~L. Chabinyc, X.~Chen, R.~E. Holmlin, H.~Jacobs, H.~Skulason, C.~D. Frisbie,
  V.~Mujica, M.~A. Ratner, M.~A. Rampi, and G.~M. Whitesides,
\newblock J.~Am.~Chem.~Soc.~ {\bf 124}, 11730 (2002).

\bibitem{choi+08sci}
S.~H. Choi, B.~Kim, and C.~D. Frisbie,
\newblock Science {\bf 320}, 1482 (2008).

\bibitem{tomf-sank02prb}
J.~Tomfohr and O.~Sankey,
\newblock Phys. Rev. B {\bf 65}, 245105 (2002).

\bibitem{peng+09jpcc}
G.~Peng, M.~Strange, K.~S. Thygesen, and M.~Mavrikakis,
\newblock J Phys Chem C {\bf 113}, 20967 (2009).

\bibitem{prod-car09prb}
E.~Prodan and R.~Car,
\newblock Phys. Rev. B {\bf 80}, 035124 (2009).

\bibitem{oeve+87jacs}
H.~OEVERING, M.~PADDONROW, M.~HEPPENER, A.~OLIVER, E.~COTSARIS, J.~VERHOEVEN,
  and N.~HUSH,
\newblock J. Am. Chem. Soc. {\bf 109}, 3258 (1987).

\bibitem{fink-hans92jacs}
H.~FINKLEA and D.~HANSHEW,
\newblock J. Am. Chem. Soc. {\bf 114}, 3173 (1992).

\bibitem{smal+95jpc}
J.~SMALLEY, S.~FELDBERG, C.~CHIDSEY, M.~LINFORD, M.~NEWTON, and Y.~LIU,
\newblock J Phys Chem-Us {\bf 99}, 13141 (1995).

\bibitem{paul+93jpc}
B.~PAULSON, K.~PRAMOD, P.~EATON, G.~CLOSS, and J.~MILLER,
\newblock J Phys Chem-Us {\bf 97}, 13042 (1993).

\bibitem{davi+98nat}
W.~Davis, W.~Svec, M.~Ratner, and M.~Wasielewski,
\newblock Nature~(London)~ {\bf 396}, 60 (1998).

\bibitem{lewi+00jacs}
F.~Lewis, T.~Wu, X.~Liu, R.~Letsinger, S.~Greenfield, S.~Miller, and
  M.~Wasielewski,
\newblock J. Am. Chem. Soc. {\bf 122}, 2889 (2000).

\bibitem{fuku-tana98ac}
K.~Fukui and K.~Tanaka,
\newblock Angew Chem Int Edit {\bf 37}, 158 (1998).

\bibitem{kell-bart99sci}
S.~O. Kelley and J.~K. Barton,
\newblock Science {\bf 283}, 375 (1999).

\bibitem{pica+03jpcm}
F.~Picaud, A.~Smogunov, A.~D. Corso, and E.~Tosatti,
\newblock J Phys-Condens Mat {\bf 15}, 3731 (2003).

\bibitem{fett-wale71book}
A.~L. Fetter and J.~D. Walecka,
\newblock {\em Quantum theory of many-particle systems},
\newblock McGraw-Hill, New York, 1971.

\bibitem{hedi65pr}
L.~Hedin,
\newblock Phys.~Rev.~ {\bf 139}, A796 (1965).

\bibitem{hedi-lund69ssp}
L.~Hedin and S.~Lundqvist,
\newblock Solid State Phys.~ {\bf 23}, 1 (1969).

\bibitem{quek+09nl}
S.~Y. Quek, H.~J. Choi, S.~G. Louie, and J.~B. Neaton,
\newblock Nano Lett {\bf 9}, 3949 (2009).

\bibitem{kumm-kron08rmp}
S.~K{\"u}mmel and L.~Kronik,
\newblock Reviews of Modern Physics {\bf 80}, 3 (2008).

\bibitem{burr+90nat}
J.~H. Burroughes, D.~D.~C. Bradley, A.~R. Brown, R.~N. Marks, K.~Mackay, R.~H.
  Friend, P.~L. Burns, and A.~B. Holmes,
\newblock Nature~(London)~ {\bf 347}, 539 (1990).

\bibitem{choi+10jacs}
S.~H. Choi, C.~Risko, M.~C.~R. Delgado, B.~Kim, J.-L. Bredas, and C.~D.
  Frisbie,
\newblock J. Am. Chem. Soc. {\bf 132}, 4358 (2010).

\bibitem{ters84prl}
J.~Tersoff,
\newblock Phys. Rev. Lett. {\bf 52}, 465 (1984).

\bibitem{kohn59pr}
W.~Kohn,
\newblock Physical Review {\bf 115}, 809 (1959).

\bibitem{he-vand01prl}
L.~He and D.~Vanderbilt,
\newblock Phys. Rev. Lett. {\bf 86}, 5341 (2001).

\bibitem{WanT}
A.~Ferretti, B.~Bonferroni, A.~Calzolari, and M.~{Buongiorno Nardelli}, 2007,
\newblock \textsc{WanT} code, {\texttt{http://www.wannier-transport.org}}.

\bibitem{ferr+07jpcm}
A.~Ferretti, A.~Calzolari, B.~Bonferroni, and R.~{Di Felice},
\newblock J.~Phys.~Condens.~Matter.~ {\bf 19}, 036215 (2007).

\bibitem{sai+05prl}
N.~Sai, M.~Zwolak, G.~Vignale, and M.~{Di Ventra},
\newblock Phys. Rev. Lett. {\bf 94}, 186810 (2005).

\bibitem{boke+07prb}
P.~Bokes, J.~Jung, and R.~W. Godby,
\newblock Phys. Rev. B {\bf 76}, 125433 (2007).

\bibitem{kurt+05prb}
S.~Kurth, G.~Stefanucci, C.~Almbladh, A.~Rubio, and E.~Gross,
\newblock Phys. Rev. B {\bf 72}, 035308 (2005).

\bibitem{thyg-rubi08prb}
K.~S. Thygesen and A.~Rubio,
\newblock Phys. Rev. B {\bf 77}, 115333 (2008).

\bibitem{vign-dive09prb}
G.~Vignale and M.~{Di Ventra},
\newblock Phys. Rev. B {\bf 79}, 014201 (2009).

\bibitem{ness+10prb}
H.~Ness, L.~K. Dash, and R.~W. Godby,
\newblock Phys. Rev. B {\bf 82}, 085426 (2010).

\bibitem{kurt+10prl}
S.~Kurth, G.~Stefanucci, E.~Khosravi, C.~Verdozzi, and E.~K.~U. Gross,
\newblock Phys. Rev. Lett. {\bf 104}, 236801 (2010).

\bibitem{ferr+05prl}
A.~Ferretti, A.~Calzolari, R.~{Di Felice}, F.~Manghi, M.~J. Caldas,
  M.~{Buongiorno Nardelli}, and E.~Molinari,
\newblock Phys.~Rev.~Lett.~ {\bf 94}, 116802 (2005).

\bibitem{ferr+05prb}
A.~Ferretti, A.~Calzolari, R.~{Di Felice}, and F.~Manghi,
\newblock Phys.~Rev.~B {\bf 72}, 125114 (2005).

\bibitem{tohe+05prl}
C.~Toher, A.~Filippetti, S.~Sanvito, and K.~Burke,
\newblock Phys. Rev. Lett. {\bf 95}, 146402 (2005).

\bibitem{ke+07jcp}
S.-H. Ke, H.~U. Baranger, and W.~Yang,
\newblock Journal of Chemical Physics {\bf 126}, 201102 (2007).

\bibitem{mera+10prb}
H.~Mera, K.~Kaasbjerg, Y.~M. Niquet, and G.~Stefanucci,
\newblock Phys. Rev. B {\bf 81}, 035110 (2010).

\bibitem{mera-niqu10prl}
H.~Mera and Y.~Niquet,
\newblock Phys. Rev. Lett. {\bf 105}, 216408 (2010).

\bibitem{lind-ratn07amat}
S.~Lindsay and M.~Ratner,
\newblock Adv. Mater {\bf 19}, 23 (2007).

\bibitem{dive09amat}
M.~D. Ventra,
\newblock Adv. Mater. {\bf 21}, 1547 (2009).

\bibitem{land70pm}
R.~Landauer,
\newblock Philos.~Mag.~ {\bf 21}, 863 (1970).

\bibitem{datt-tian97prb}
S.~Datta and W.~Tian,
\newblock Phys. Rev. B {\bf 55}, 1914 (1997).

\bibitem{rang+11prb}
T.~Rangel, A.~Ferretti, P.~E. Trevisanutto, V.~Olevano, and G.~M. Rignanese,
\newblock Phys. Rev. B {\bf 84}, 045426 (2011).

\bibitem{stra+11prb}
M.~Strange, C.~Rostgaard, H.~H{\"a}kkinen, and K.~Thygesen,
\newblock Phys. Rev. B {\bf 83}, 115108 (2011).

\bibitem{dara+07prb}
P.~Darancet, A.~Ferretti, D.~Mayou, and V.~Olevano,
\newblock Phys.~Rev.~B {\bf 75}, 075102 (2007).

\bibitem{thyg-rubi07jcp}
K.~S. Thygesen and A.~Rubio,
\newblock J.~Chem.~Phys.~ {\bf 126}, 091101 (2007).

\bibitem{ceho+08prb}
A.~Cehovin, H.~Mera, J.~H. Jensen, K.~Stokbro, and T.~B. Pedersen,
\newblock Phys. Rev. B {\bf 77}, 195432 (2008).

\bibitem{mowb+08jcp}
D.~J. Mowbray, G.~Jones, and K.~S. Thygesen,
\newblock Journal of Chemical Physics {\bf 128}, 111103 (2008).

\bibitem{quek+07nl}
S.~Y. Quek, L.~Venkataraman, H.~J. Choi, S.~G. Louie, M.~S. Hybertsen, and
  J.~B. Neaton,
\newblock Nano~Lett.~ {\bf 7}, 3477 (2007).

\bibitem{step+94jpc}
P.~Stephens, F.~Devlin, C.~CHABALOWSKI, and M.~Frisch,
\newblock J Phys Chem-Us {\bf 98}, 11623 (1994).

\bibitem{perd+96jcp}
J.~Perdew, M.~Ernzerhof, and K.~Burke,
\newblock Journal of Chemical Physics {\bf 105}, 9982 (1996).

\bibitem{musc+01cpl}
J.~Muscat, A.~Wander, and N.~Harrison,
\newblock Chemical Physics Letters {\bf 342}, 397 (2001).

\bibitem{blas+11prb}
X.~Blase, C.~Attaccalite, and V.~Olevano,
\newblock Phys. Rev. B {\bf 83}, 115103 (2011).

\bibitem{jain+11prl}
M.~Jain, J.~Chelikowsky, and S.~Louie,
\newblock Phys. Rev. Lett. {\bf 107}, 216806 (2011).

\bibitem{refa+11prb}
S.~Refaely-Abramson, R.~Baer, and L.~Kronik,
\newblock Phys. Rev. B {\bf 84}, 075144 (2011).

\bibitem{rost+10prb}
C.~Rostgaard, K.~W. Jacobsen, and K.~S. Thygesen,
\newblock Phys. Rev. B {\bf 81}, 085103 (2010).

\bibitem{onid+02rmp}
G.~Onida, L.~Reining, and A.~Rubio,
\newblock Rev.~Mod.~Phys.~ {\bf 74}, 601 (2002).

\bibitem{hybe-loui86prb1}
M.~Hybertsen and S.~Louie,
\newblock Phys. Rev. B {\bf 34}, 5390 (1986).

\bibitem{hybe-loui86prb}
M.~S. Hybertsen and S.~G. Louie,
\newblock Phys.~Rev.~B {\bf 34}, 2920 (1986).

\bibitem{arya-gunn98rpp}
F.~Aryasetiawan and O.~Gunnarsson,
\newblock Rep.~Prog.~Phys.~ {\bf 61}, 237 (1998).

\bibitem{godb-need89prl}
R.~Godby and R.~NEEDS,
\newblock Phys. Rev. Lett. {\bf 62}, 1169 (1989).

\bibitem{heyd+03jcp}
J.~Heyd, G.~Scuseria, and M.~Ernzerhof,
\newblock J. Chem. Phys. {\bf 118}, 8207 (2003).

\bibitem{perd+07pra}
J.~Perdew, A.~Ruzsinszky, G.~Csonka, O.~Vydrov, G.~Scuseria, V.~Staroverov, and
  J.~Tao,
\newblock Phys. Rev. A {\bf 76}, 040501 (2007).

\bibitem{marq+11prb}
M.~Marques, J.~Vidal, M.~Oliveira, L.~Reining, and S.~Botti,
\newblock Phys. Rev. B {\bf 83}, 035119 (2011).

\bibitem{bloc+94prb}
P.~E. Bl{\"o}chl, O.~Jepsen, and O.~K. Andersen,
\newblock Phys.~Rev.~B {\bf 49}, 16223 (1994).

\bibitem{yate+07prb}
J.~Yates, X.~Wang, D.~Vanderbilt, and I.~Souza,
\newblock Phys. Rev. B {\bf 75}, 195121 (2007).

\bibitem{marz-vand97prb}
N.~Marzari and D.~Vanderbilt,
\newblock Phys.~Rev.~B {\bf 56}, 12847 (1997).

\bibitem{souz+01prb}
I.~Souza, N.~Marzari, and D.~Vanderbilt,
\newblock Phys.~Rev.~B {\bf 65}, 035109 (2001).

\bibitem{hama-vand09prb}
D.~R. Hamann and D.~Vanderbilt,
\newblock Phys. Rev. B {\bf 79}, 045109 (2009).

\bibitem{CRYSTAL09}
R.~Dovesi, V.~R. Saunders, C.~Roetti, R.~Orlando, C.~M. Zicovich-Wilson,
  F.~Pascale, B.~Civalleri, K.~Doll, N.~M. Harrison, I.~J. Bush, P.~{D'Arco},
  and M.~Llunell,
\newblock {\em {\sc CRYSTAL09} User's Manual},
\newblock Universit\`a di Torino, Torino, 2009.

\bibitem{note_ortho_crystal}
The local non-orthogonal basis set of CRYSTAL is orthogonalized ($S'=Id;
  H'=S^{-1/2}\, H \,S^{-1/2}$) to improve the numerical accuracy of the complex
  band structure interpolation. Only in the numerically critical case of
  poly-ethylene (PE), the CRYSTAL basis is not orthogonalized and a real space
  cutoff of the Hamiltonian is applied [$H(R) = 0 $ for $R \geq 3$ unit cells].
  For the same system, a real space cutoff has been applied to $H(R)$ also when
  using Wannier functions as a basis (for $R \geq 4$ unit cells).

\bibitem{mart-buss09cpc}
L.~Martin-Samos and G.~Bussi,
\newblock Computer Physics Communications {\bf 180}, 1416 (2009).

\bibitem{gian+09jpcm}
P.~Giannozzi, S.~Baroni, N.~Bonini, M.~Calandra, R.~Car, C.~Cavazzoni,
  D.~Ceresoli, G.~L. Chiarotti, M.~Cococcioni, I.~Dabo, A.~D. Corso,
  S.~de~Gironcoli, S.~Fabris, G.~Fratesi, R.~Gebauer, U.~Gerstmann,
  C.~Gougoussis, A.~Kokalj, M.~Lazzeri, L.~Martin-Samos, N.~Marzari, F.~Mauri,
  R.~Mazzarello, S.~Paolini, A.~Pasquarello, L.~Paulatto, C.~Sbraccia,
  S.~Scandolo, G.~Sclauzero, A.~P. Seitsonen, A.~Smogunov, P.~Umari, and R.~M.
  Wentzcovitch,
\newblock J Phys-Condens Mat {\bf 21}, 395502 (2009).

\bibitem{vanf+02prl}
M.~V. Faassen and P.~de~Boeij,
\newblock Journal of Chemical Physics {\bf 120}, 8353 (2004).

\bibitem{kirt+95jcp}
B.~Kirtman, J.~Toto, K.~Robins, and M.~Hasan,
\newblock J. Chem. Phys. {\bf 102}, 5350 (1995).

\bibitem{rodr-lars95jcp}
L.~Rodr{\'\i}guez-Monge and S.~Larsson,
\newblock J.~Chem.~Phys.~ {\bf 102}, 7106 (1995).

\bibitem{rohl+01sm}
M.~Rohlfing, M.~Tiago, and S.~Louie,
\newblock Synthetic Met {\bf 116}, 101 (2001).

\bibitem{tiag+04prb}
M.~Tiago, M.~Rohlfing, and S.~Louie,
\newblock Phys. Rev. B {\bf 70}, 193204 (2004).

\bibitem{rohl-loui99prl}
M.~Rohlfing and S.~G. Louie,
\newblock Phys.~Rev.~Lett.~ {\bf 82}, 1959 (1999).

\bibitem{yann-clar83prl}
C.~YANNONI and T.~CLARKE,
\newblock Phys. Rev. Lett. {\bf 51}, 1191 (1983).

\bibitem{zhu+92ssc}
Q.~ZHU, J.~FISCHER, R.~ZUZOK, and S.~Roth,
\newblock Solid State Commun {\bf 83}, 179 (1992).

\bibitem{pusc-ambr02prl}
P.~Puschnig and C.~Ambrosch-Draxl,
\newblock Phys.~Rev.~Lett.~ {\bf 89}, 056405 (2002).

\bibitem{prod-car08nl}
E.~Prodan and R.~Car,
\newblock Nano Lett {\bf 8}, 1771 (2008).

\bibitem{stra-thyg11bjn}
M.~Strange and K.~S. Thygesen,
\newblock Beilstein J. Nanotechnol. {\bf 2}, 746 (2011).

\bibitem{note_COHSEX}
In the case of COHSEX, the screening $W$ is computed at the LDA level ($W_0$)
  and it is not updated during self-consistency.

\bibitem{varsano_phd}
D. Varsano, {\it First principles description of response functions in low
  dimensional systems}, Phd thesis (2006).

\bibitem{tamb+11prb}
I.~Tamblyn, P.~Darancet, S.~Y. Quek, S.~Bonev, and J.~Neaton,
\newblock Phys. Rev. B {\bf 84}, 201402 (2011).

\bibitem{tiag+08jcp}
M.~L. Tiago, P.~R.~C. Kent, R.~Q. Hood, and F.~A. Reboredo,
\newblock Journal of Chemical Physics {\bf 129}, 084311 (2008).

\bibitem{kaas-thyg10prb}
K.~Kaasbjerg and K.~S. Thygesen,
\newblock Phys. Rev. B {\bf 81}, 085102 (2010).

\bibitem{fabe+11prb}
C.~Faber, C.~Attaccalite, V.~Olevano, E.~Runge, and X.~Blase,
\newblock Phys. Rev. B {\bf 83}, 115123 (2011).

\bibitem{rina+01prb}
R.~Rinaldi, R.~Cingolani, K.~M. Jones, A.~A. Baski, H.~Morkoc, A.~D. Carlo,
  J.~Widany, F.~D. Sala, and P.~Lugli,
\newblock Phys.~Rev.~B {\bf 63}, 075311 (2001).

\bibitem{keme+04prb}
M.~Kemerink, S.~Alvarado, P.~Muller, P.~Koenraad, H.~Salemink, J.~Wolter, and
  R.~Janssen,
\newblock Phys. Rev. B {\bf 70}, 045202 (2004).

\bibitem{ters84prb}
J.~Tersoff,
\newblock Phys. Rev. B {\bf 30}, 4874 (1984).

\bibitem{venk+06nl}
L.~Venkataraman, J.~Klare, I.~Tam, C.~Nuckolls, M.~Hybertsen, and
  M.~Steigerwald,
\newblock Nano Lett {\bf 6}, 458 (2006).

\bibitem{park+09jacs}
Y.~S. Park, J.~R. Widawsky, M.~Kamenetska, M.~L. Steigerwald, M.~S. Hybertsen,
  C.~Nuckolls, and L.~Venkataraman,
\newblock J. Am. Chem. Soc. {\bf 131}, 10820 (2009).

\bibitem{hybe+08jpcm}
M.~Hybertsen, L.~Venkataraman, J.~Klare, A.~Whalley, M.~Steigerwald, and
  C.~Nuckolls,
\newblock J. Phys. Condens. Matter {\bf 20}, 374115 (2008).

\bibitem{wold+02jpcb}
D.~Wold, R.~Haag, M.~Rampi, and C.~Frisbie,
\newblock J Phys Chem B {\bf 106}, 2813 (2002).

\bibitem{xu-tao03sci}
B.~Xu and N.~Tao,
\newblock Science {\bf 301}, 1221 (2003).

\bibitem{kaun+03prb}
C.~Kaun, B.~Larade, and H.~Guo,
\newblock Phys. Rev. B {\bf 67}, 121411 (2003).

\bibitem{venk+06nat}
L.~Venkataraman, J.~E. Klare, C.~Nuckolls, M.~S. Hybertsen, and M.~L.
  Steigerwald,
\newblock Nature {\bf 442}, 904 (2006).

\bibitem{he+05jacs}
J.~He, F.~Chen, J.~Li, O.~Sankey, Y.~Terazono, C.~Herrero, D.~Gust, T.~Moore,
  A.~Moore, and S.~Lindsay,
\newblock J. Am. Chem. Soc. {\bf 127}, 1384 (2005).

\bibitem{meis+11nl}
J.~S. Meisner, M.~Kamenetska, M.~Krikorian, M.~L. Steigerwald, L.~Venkataraman,
  and C.~Nuckolls,
\newblock Nano Lett {\bf 11}, 1575 (2011).

\bibitem{note-details-crystal}
The Coulomb and exchange series are summed directly and truncated using overlap
  criteria with thresholds of 10$^{-10}$, 10$^{-10}$, 10$^{-10}$, 10$^{-10}$
  and 10$^{-20}$ as described previously;~\cite{Pisani_1988,CRYSTAL06} the
  choice of these very severe thresholds has been dictated by the high accuracy
  required by the complex band structure calculations.

\bibitem{mart+11tobe}
To this aim, electrostatic interactions were cut off in real space for
  distances larger than the Wigner-Seitz supercell corresponding to the adopted
  $\mathbf{k}$-point mesh (L. Martin-Samos, A. Ferretti, G. Bussi, to be
  published).

\bibitem{mass+93prb}
S.~Massidda, M.~Posternak, and A.~Baldereschi,
\newblock Phys. Rev. B {\bf 48}, 5058 (1993).

\bibitem{mari+09cpc}
A.~Marini, C.~Hogan, M.~Gr{\"u}ning, and D.~Varsano,
\newblock Computer physics communications {\bf 180}, 1392 (2009).

\bibitem{isma06prb}
S.~Ismail-Beigi,
\newblock Phys. Rev. B {\bf 73}, 233103 (2006).

\bibitem{li-dabo11prb}
Y.~Li and I.~Dabo,
\newblock Phys. Rev. B {\bf 84}, 155127 (2011).

\bibitem{rozz+06prb}
C.~A. Rozzi, D.~Varsano, A.~Marini, E.~K.~U. Gross, and A.~Rubio,
\newblock Phys. Rev. B {\bf 73}, 205119 (2006).

\bibitem{Pisani_1988}
C.~Pisani, R.~Dovesi, and C.~Roetti,
\newblock {\em Hartree-Fock ab initio Treatment of Crystalline Systems},
  volume~48 of {\em Lecture Notes in Chemistry},
\newblock Springer Verlag, Heidelberg, 1988.

\bibitem{CRYSTAL06}
R.~Dovesi, V.~R. Saunders, C.~Roetti, R.~Orlando, C.~M. Zicovich-Wilson,
  F.~Pascale, B.~Civalleri, K.~Doll, N.~M. Harrison, I.~J. Bush, P.~{D'Arco},
  and M.~Llunell,
\newblock {\em {\sc CRYSTAL06} User's Manual},
\newblock Universit\`a di Torino, Torino, 2006.

\end{thebibliography}
\end{document}